\documentclass[twocolumn]{aastex62}

\usepackage{amsmath, amsthm, amssymb, commath, graphicx, bbold, endnotes, graphicx, subfigure, multirow, comment}

\newcommand\fitparams{\boldsymbol{\theta}}
\newcommand\visfunc{\boldsymbol{\zeta}}
\newcommand\freqparams{\boldsymbol{\gamma}}
\newcommand\thermalcov{\boldsymbol{\mathsf{C}}_\text{T}}
\newcommand\modelcov{\boldsymbol{\mathsf{C}}_\text{M}}
\newcommand\modelcorr{\boldsymbol{\mathsf{C}}_\text{R}}
\newcommand\data{\boldsymbol{v}}
\newcommand\fisherinfo{\boldsymbol{\mathsf{I}}}
\newcommand\gains{\boldsymbol{g}}
\newcommand\gainsmat{\boldsymbol{\mathsf{G}}}
\newcommand\fitparamsu{\boldsymbol{u}}
\newcommand\modelvals{\boldsymbol{m}}
\newcommand\thermalvar{\boldsymbol{\sigma}_\text{T}^2}
\newcommand\modelvar{\boldsymbol{\sigma}_\text{M}^2}
\newcommand\matrixa{\boldsymbol{\mathsf{A}}}
\newcommand\uvcoord{\boldsymbol{x}}
\newcommand\beamvec{\boldsymbol{B}}
\newcommand\groundcoord{\boldsymbol{r}}
\newcommand\antres{\boldsymbol{b}}
\newcommand\beammat{\boldsymbol{\mathsf{B}}}
\newcommand\jonesmat{\boldsymbol{\mathsf{J}}}
\newcommand\coherency{\boldsymbol{\mathsf{S}}}
\newcommand\electricfield{\boldsymbol{E}}
\newcommand\vismat{\boldsymbol{\mathsf{V}}}
\newcommand\negloglikelihood{L}
\newcommand\gainamp{A}
\newcommand\unnormgains{\boldsymbol{h}}

\begin{document}
\title{A Unified Calibration Framework for 21 cm Cosmology}
\submitjournal{Monthly Notices of the Royal Astronomical Society}
\accepted{March 2, 2021}

\author{Ruby Byrne}
\affiliation{Physics Department, University of Washington, Seattle, WA, USA}

\author{Miguel F. Morales}
\affiliation{Physics Department, University of Washington, Seattle, WA, USA}

\author{Bryna J. Hazelton}
\affiliation{Physics Department, University of Washington, Seattle, WA, USA}
\affiliation{eScience Institute, University of Washington, Seattle, WA, USA}

\author{Michael Wilensky}
\affiliation{Physics Department, University of Washington, Seattle, WA, USA}

\begin{abstract}
    Calibration precision is currently a limiting systematic in 21 cm cosmology experiments. While there are innumerable calibration approaches, most can be categorized as either `sky-based,' relying on an extremely accurate model of astronomical foreground emission, or `redundant,' requiring a precisely regular array with near-identical antenna response patterns. Both of these classes of calibration are inflexible to the realities of interferometric measurement. In practice, errors in the foreground model, antenna position offsets, and beam response inhomogeneities degrade calibration performance and contaminate the cosmological signal. Here we show that sky-based and redundant calibration can be unified into a highly general and physically motivated calibration framework based on a Bayesian statistical formalism. Our new framework includes sky-based and redundant calibration as special cases but can additionally support relaxing the rigid assumptions implicit in those approaches. We present simulation results demonstrating that, in a simple case, working in an intermediate regime between sky-based and redundant calibration improves calibration performance. Our framework is highly general and encompasses novel calibration approaches including techniques for calibrating compact non-redundant arrays, calibrating to incomplete sky models, and constraining calibration solutions across frequency.
\end{abstract}


\section{Introduction}

Measurement of the 21 cm cosmological power spectrum would constrain models of the Epoch of Reionization (EoR), Dark Ages, and Dark Energy. These measurements are contaminated by astrophysical foreground emission that is 4-5 orders-of-magnitude brighter than the cosmological signal. Separating the cosmological and foreground signals requires extremely precise instrumental calibration. As a result, development and characterization of precision calibration techniques for 21 cm cosmology has become an active area of research \citep{Wieringa1992, Pen2009, Liu2010, Kazemi2011, Kazemi2013, Grobler2014, Newburgh2014, Salvini2014, Zheng2014, Yatawatta2015, Barry2016, Berger2016, Dillon2016,  Ewall-Wice2017, Grobler2016, Patil2016, Weeren2016, Wijnholds2016, Ollier2017, Dillon2018, Grobler2018, Joseph2018, Li2018, Tasse2018, Byrne2019, Kohn2019, Li2019, Orosz2019, Albert2020, Dillon2020, Joseph2020, Kern2019, Mertens2020, Sob2020}.

In general, calibration approaches for cosmological 21 cm power spectrum measurements can be categorized as one of two types. `Sky-based' calibration uses models of the sky and instrument to simulate data. Calibration then consists of fitting measurements to the simulation. In contrast, `redundant calibration' relies on highly regular arrays with many redundant baselines measuring the same sky signal \citep{Wieringa1992, Liu2010, Dillon2016, Dillon2018, Grobler2018, Kern2019,Dillon2020}. Calibration then fits redundant measurements to each other, aiming for consistency between the measurements. Many diverse approaches fall into one of these two classes of calibration.

As the field of 21 cm cosmology pushes the limits of precision interferometric calibration it has become increasingly important that calibration frameworks mitigate error while capturing all instrumental systematics. Sky-based calibration approaches assume very good \textit{a priori} models of the sky and instrument; model errors quickly degrade calibration and can preclude detection of the cosmological signal \citep{Barry2016, Ewall-Wice2017}. Redundant calibration is likewise vulnerable to sky model errors  \citep{Byrne2019} and experiences further calibration errors from instrument non-redundancies. Antenna position offsets and beam response inhomogeneities break redundant calibration's assumption of baseline redundancy and produce calibration errors \citep{Joseph2018, Li2018, Orosz2019}. The field requires novel calibration approaches that are resilient to sky and instrument model errors, can capture array non-redundancies, and mitigate contamination of the cosmological signal.

Combined calibration approaches integrate aspects of sky-based and redundant calibration. The Murchison Widefield Array (MWA) Phase II is a hybrid array that incorporates regular hexagonal sub-arrays within an otherwise pseudo-random array \citep{Wayth2018}. \citealt{Li2018}, \citealt{Li2019}, and \citealt{Zhang2020} combine redundant calibration of the sub-arrays with sky-based calibration for the remainder of the array. \citealt{Sievers2017} proposes a calibration algorithm that relaxes redundancy requirements for redundant calibration of a regular array. This paper expands on these ideas to present a fully unified calibration framework that is statistically rigorous, highly flexible, and physically-motivated. 

In \S\ref{s:background} we introduce a calibration formalism based on a Bayesian statistical approach. In \S\ref{s:trad_cal} we re-derive simple sky-based and redundant calibration frameworks with a focus on delineating the implicit assumptions of those approaches. In \S\ref{s:general_framework} we present a novel calibration framework that unifies redundant and sky-based calibration, and \S\ref{s:sim_results} presents simulation results that explore some of the implications of that framework. Next, we discuss extensions to the framework that allow for calibration across frequencies (\S\ref{s:freq_cal}) and fully-polarized calibration (\S\ref{s:pol_cal}). Table \ref{t:var_table} presents a summary of variables and expressions used throughout the paper.

\section{Statistical Formalism}
\label{s:background}

In this section we present statistical underpinnings of the calibration problem. We use Bayes' theorem to formulate a likelihood function with a statistical prior. Calibration consists of maximizing this likelihood function to calculate the instrumental response.

\begin{table*}
\centering
\begin{tabular}{ |c|p{10cm}| } 
 \hline
 \textbf{Variable or Expression} & \textbf{Definition} \\ 
 \hline
 $\data$ & data (for interferometric data, these are the measured visibilities) \\ 
 \hline
 $\langle \data \rangle$ & expectation value of the data \\ 
 \hline
 $\fitparams$ & tunable calibration parameters \\ 
 \hline
 $\hat{\fitparams}$ & maximum-likelihood calibration parameters \\ 
 \hline
 $\visfunc(\fitparams)$ & function that maps the calibration parameters to the model of the data \\
 \hline
 $P(\fitparams | \data)$ & likelihood function, equal to the probability that $\fitparams$ parameterizes the model given the data $\data$ \\
 \hline
 $P(\data | \fitparams)$ & probability that the model parameterized by $\fitparams$ produces data $\data$ \\
 \hline
 $P(\fitparams)$ & prior probability of $\fitparams$ \\
 \hline
 $P(\data)$ & marginalized probability of $\data$ \\
 \hline
 $\negloglikelihood(\fitparams)$ & negative log-likelihood; for a Gaussian likelihood function $\negloglikelihood(\fitparams)$ is also called the chi-squared (or $\chi^2$) \\
 \hline
 $\fisherinfo(\fitparams)$ & Fisher information matrix \\
 \hline
 $\gains$ & antenna gains; these are an example of calibration parameters $\fitparams$ \\
 \hline
 $\gainsmat(\gains)$ & antenna gains written as a matrix; diagonal matrix with elements $\gains_a \gains_b^*$ \\
 \hline
 $\fitparamsu$ & additional calibration parameters that are not $\gains$ \\
 \hline
 $\modelvals$ & the \textit{a priori} estimator of $\fitparamsu$ \\
 \hline
 $\thermalcov$ & thermal covariance matrix of the data $\data$ \\
 \hline
 $\thermalvar$ & elements of $\thermalcov$ when $\thermalcov$ is diagonal \\
 \hline
 $\modelcov$ & covariance matrix of the calibration parameters $\fitparamsu$ \\
 \hline
 $\modelvar$ & elements of $\modelcov$ when $\modelcov$ is diagonal \\
 \hline
 $\matrixa$ & matrix that maps calibration parameters $\fitparamsu$ to the data \\
 \hline
 $N_\text{ant}$ & number of antennas \\
 \hline
 $N_\text{red}$ & number of redundant baseline sets \\
 \hline
 $\uvcoord$ & \textit{uv} coordinate vector \\
 \hline
 $\beamvec_j(\uvcoord)$ & \textit{uv} response of baseline $j$, assuming a continuous \textit{uv} plane \\
 \hline
 $\beammat$ & matrix that encodes the \textit{uv} responses of baselines \\
 \hline
 $f$ & frequency \\
 \hline
 $\freqparams_a$ & calibration parameters that parameterize the gain of antenna $a$ across frequency; these are an example of calibration parameters $\fitparams$ \\
 \hline
 $\vismat_{ab}$ & $2\times2$ visiblity matrix representing the 4 polarizations measured from baseline $\{a,b\}$ \\
 \hline
 $\jonesmat_{ak}$ & $2\times2$ Jones matrix for antenna $a$ at sky location $k$ \\
 \hline
 $\coherency_k$ & $2\times2$ sky coherency matrix at location $k$ \\
 \hline
 $\electricfield_k$ & polarized electric field vector of radiation emanating from sky location $k$ \\
 \hline
 $\eta$ & delay, Fourier dual of frequency $f$ \\
 \hline
\end{tabular}
\caption{A summary of variables and expressions used throughout the paper. Non-bold italicized variables represent scalars (e.g. $f$), bold italicized variables represent vectors (e.g. $\boldsymbol{v}$), and bold sans-serif variables represent matrices (e.g. $\boldsymbol{\mathsf{A}}$). Italicized subscripts represent indices (e.g. $\boldsymbol{v}_j$ is the $j$-th element of the vector $\boldsymbol{v}$) and non-italic subscripts are simply descriptive (e.g. $\boldsymbol{u}_\text{orig}$). $\log$ represents the natural logarithm and we denote the identity matrix $\mathbb{1}$.}
\label{t:var_table}
\end{table*}

\subsection{Bayesian Statement of the Calibration Problem}

Calibration can be interpreted as a model fitting problem that fits tunable parameters to data. The fitting procedure asks, `what are the most likely calibration parameters given the data?' If we define our visibility data as $\data$ and our tunable calibration parameters as $\fitparams$, we can write the probability of $\fitparams$ given $\data$ as $P(\fitparams|\data)$. This quantity is also called the `likelihood function.' We can use the likelihood function to calculate maximum likelihood parameters $\hat{\fitparams}$ that maximize $P(\fitparams|\data)$.

As in most model fitting problems, there is no single `right answer' for the form of the likelihood. One must construct a model that appropriately represents the data and mitigates systematics that contaminate the measurement. As precision calibration is crucial to the success of 21cm cosmology, choosing a well-motivated form for the likelihood function is extremely important. This paper describes a general calibration framework and explores different physically motivated likelihood functions that could improve calibration performance for 21cm cosmology.

From Bayes' theorem we can write the likelihood function as
\begin{equation}
    P(\fitparams|\data) = \frac{P(\data|\fitparams) P(\fitparams)}{P(\data)}
\end{equation}
where $P(\data|\fitparams)$ is the probability of the data $\data$ given model parameters $\fitparams$. $P(\fitparams)$ is the prior probability, i.e.\ the probability of $\fitparams$ independent of the data. $P(\data)$ is the marginal likelihood of $\data$ and is constant across all models. We can therefore simplify the problem by considering a proportionality
\begin{equation}
    P(\fitparams|\data) \propto P(\data|\fitparams) P(\fitparams)
\label{eq:likelihood}
\end{equation}
and maximizing the right hand side. 

Maximizing the likelihood function $P(\fitparams|\data)$ is equivalent to minimizing the negative log-likelihood. Calibration algorithms typical perform the latter procedure rather than explicitly maximizing $P(\fitparams|\data)$. We describe the negative log-likelihood as
\begin{equation}
    \negloglikelihood(\fitparams) = - C_1 \log[{P(\fitparams|\data)}] + C_2,
\label{eq:log_likelihood}
\end{equation}
where $C_1$ is an arbitrary positive constant and $C_2$ is an arbitrary constant of either sign. Neither constant affects the result $\hat{\fitparams}$ achieved by minimizing $\negloglikelihood(\fitparams)$, and we therefore choose $C_1$ and $C_2$ to simplify the form of $\negloglikelihood(\fitparams)$. For the case that ${P(\fitparams|\data)}$ takes the form of a Gaussian distribution, the negative log-likelihood is equivalent to the least-squares cost function, also called the chi-squared and denoted $\chi^2$. For the sake of generality we do not assume that the likelihood is Gaussian. $\negloglikelihood(\fitparams)$ can, but does not necessarily, take the form of a chi-squared.

This formulation of the likelihood is completely general and $\fitparams$ can take any form. For example, traditional direction-independent calibration defines $\fitparams$ to be per-antenna gains; direction-dependent calibration allows these gains to assume different values at different positions on the sky. Other calibration approaches such as redundant calibration may introduce additional tunable model parameters that parameterize uncertainties on the sky model. As with the gains, calibration tunes these additional parameters to the values that maximize the likelihood function.

$\fitparams$ models the data according to some general function $\visfunc(\fitparams)$. Our calibration model assumes that this function can reconstruct the expectation value of the data for some values of the tunable parameters:
\begin{equation}
    \langle \data \rangle = \visfunc(\hat{\fitparams}),
\end{equation}
where $\langle \data \rangle$ is the expectation value of the data.

A specific calibration approach is defined by the following choices:
\begin{enumerate}
    \item A choice of the tunable model parameters $\fitparams$. These parameterize the instrument model and could parameterize the sky model as well.
    \item A choice of the form of the visibility model as a function of the tunable parameters $\visfunc(\fitparams)$.
    \item A choice of the form of $P(\data|\fitparams)$ that describes the noise properties of the data.
    \item A choice of the prior on the tunable parameters $P(\fitparams)$. This could be a flat prior, such that $P(\fitparams)$ is a constant, or it could have a functional form where $\hat{\fitparams}$ favors an expectation value.
\end{enumerate}

Once these choices are defined, calibration consists of fitting maximum-likelihood parameters $\hat{\fitparams}$. A final choice determines how these parameters are applied to the data to produce calibrated results. In some calibration frameworks only a subset of the fitted parameters get applied to the data. The rest of the fitted parameters are internal to the calibration optimization procedure. In general, calibration parameters related to the instrument response are involved in transforming uncalibrated data into calibrated data. Calibration parameters related to the sky model are internal to calibration.

All calibration frameworks described in this paper assume that the thermal noise properties of the data are Gaussian. The probability distribution $P(\data|\fitparams)$ from item (3) above therefore takes the form of a Gaussian distribution. While this is the most common choice, other calibration approaches relax the assumption of Gaussianity \citep{Kazemi2013, Ollier2017, Sob2020}. The calibration frameworks in this paper can be easily extended to non-Gaussian distributions.

\subsection{Quantifying Constraint of Calibration Solutions}

A good parameterization of the model leads to highly constrained tunable parameters. The constraint on the tunable calibration parameters $\fitparams$ is quantified by the Fisher information.

The Fisher information matrix $\fisherinfo$ is given by the negative Hessian of the log-likelihood:
\begin{equation}
    \fisherinfo_{jk}(\fitparams) = - \frac{\partial^2 \log[P(\fitparams|\data)]}{\partial \fitparams_j \partial \fitparams_k}.
\end{equation}
From Equation \ref{eq:log_likelihood} we can also express the Fisher information in terms of the negative log-likelihood, $\negloglikelihood(\fitparams)$:
\begin{equation}
    \fisherinfo_{jk}(\fitparams) = \frac{\partial^2 \negloglikelihood(\fitparams)}{\partial \fitparams_j \partial \fitparams_k}.
\end{equation}

One can interpret the Fisher information as a measure of the curvature of the $\negloglikelihood(\fitparams)$ hypersurface. At the maximum-likelihood point, where $\fitparams = \hat{\fitparams}$, $\negloglikelihood(\fitparams)$ experiences a minimum. We therefore expect that all elements of the Fisher information matrix are non-negative:
\begin{equation}
    \fisherinfo_{jk}(\hat{\fitparams}) \ge 0. 
\end{equation}
Large values of this Fisher information indicate that $\negloglikelihood(\fitparams)$ experiences a sharp minimum and the tunable parameters are highly constrained. Conversely, small Fisher information values mean the model is relatively agnostic to the parameter values and that the tunable parameters are not well-constrained.

If $\fisherinfo_{jk}(\hat{\fitparams})=0$ for all $k$ then the parameter $\fitparams_j$ is completely unconstrained by the model. It follows that the calibration solutions are degenerate. $\negloglikelihood(\fitparams)$ does not experience a unique minimum but is rather minimized for any value of $\fitparams_j$. In this case, $\fitparams_j$ is a degenerate parameter.

It is not always evident when calibration solutions are degenerate. Degenerate parameters can consist of linear combinations of the tunable parameters $\fitparams$. We can calculate the number of degenerate parameters by taking the rank of the Fisher information matrix at the maximum-likelihood point. The calibration solutions are degenerate if $\fisherinfo(\hat{\fitparams})$ is singular. The number of degenerate parameters is equal to the dimensionality of the null space of $\fisherinfo(\hat{\fitparams})$; the degenerate parameters are the eigenvectors that span the null space.

Calibration degeneracies are a major challenge in 21cm cosmology. All degeneracies must be constrained to yield physical calibration solutions. Often, calibration consists of two stages where an initial calibration framework yields degenerate solutions and second step constrains those degeneracies \citep{Liu2010, Zheng2014, Byrne2019, Kern2019}.

\section{Traditional Direction-Independent Calibration Approaches}
\label{s:trad_cal}

In this section we describe two simple direction-independent calibration frameworks, which we call `sky-based calibration' and `redundant calibration.' We explain how these calibration frameworks connect with the statistical approach described in \S\ref{s:background}.

Both these calibration frameworks assume that the likelihood function is separable in frequency, time, and polarization. This means that each frequency channel, observation interval, and polarization mode can be calibrated independently, and we therefore omit explicit frequency, time, and polarization dependence. For a discussion of frequency calibration see \S\ref{s:freq_cal}; for a discussion of polarized calibration see \S\ref{s:pol_cal}. We assume per-time calibration throughout the paper.

\subsection{Sky-Based Calibration}
\label{s:sky_cal}

Traditional sky-based calibration makes the following choices:
\begin{enumerate}
    \item It parameterizes the tunable instrument calibration parameters as a complex gain per antenna (and implicitly per polarization, frequency, and time step). $\fitparams = \gains$ where $\gains$ has length equal to the number of antennas.
    \item It models the data as 
    \begin{equation}
        \visfunc_{ab}=\gains_a \gains_b^* \modelvals_{ab}
    \label{eq:vis_model_sky_cal}
    \end{equation}
    where indices $a$ and $b$ index antennas. The combined indices $ab$ indicate the index of the visibility formed by correlating signals from antennas $a$ and $b$. $\modelvals$ are model visibilities developed with a sky model and instrument simulator. The sky model typically consists of a point-source catalog, sometimes in conjunction with a diffuse foreground emission map.
    \item It describes $P(\data|\fitparams)$ as a Gaussian probability distribution:
    \begin{equation}
        P(\data|\fitparams) \propto e^{-\frac{1}{2} [\data - \visfunc(\fitparams)]^\dag \thermalcov^{-1} [\data - \visfunc(\fitparams)]},
    \end{equation}
    where the $\dag$ symbol denotes the conjugate transpose. Here $\thermalcov$ is the thermal covariance matrix, given by
    \begin{equation}
        \thermalcov = \langle (\data-\langle \data \rangle)(\data-\langle \data \rangle)^\dag \rangle.
    \end{equation}
    Furthermore, it assumes the visibilities are independent such that $\thermalcov$ is diagonal. The probability function therefore takes the form
    \begin{equation}
        P(\data|\fitparams) \propto e^{-\frac{1}{2} \sum_j \frac{1}{\thermalvar{}_{,j}} \left|\data_j - \visfunc_j(\fitparams) \right|^2}
    \label{eq:independent_gaussian}
    \end{equation}
    where $\thermalvar{}_{,j}$ are the diagonal elements of $\thermalcov$.
    \item It uses a flat prior such that $P(\fitparams)=P(\gains)$ is a constant.
\end{enumerate}

With these choices we can write the likelihood function for traditional sky-based calibration as
\begin{equation}
    P(\gains|\data) \propto e^{-\frac{1}{2} \sum_{ab} \frac{1}{\thermalvar{}_{,ab}} \left|\data_{ab} - \gains_a \gains_b^* \modelvals_{ab}\right|^2}.
\label{eq:sky_cal_likelihood}
\end{equation}
Maximizing this likelihood function is equivalent to minimizing the negative log-likelihood given by\footnote{Note that this quantity is actually twice the negative log-likelihood: $\negloglikelihood(\gains) = -2 \log[{P(\gains|\data)}]$. However, per Equation \ref{eq:log_likelihood} the coefficient is an arbitrary constant. For conciseness, throughout the paper we will refer to this quantity as the `negative log-likelihood.'}
\begin{equation}
    \negloglikelihood(\gains) = \sum_{ab} \frac{1}{\thermalvar{}_{,ab}} \left|\data_{ab} - \gains_a \gains_b^* \modelvals_{ab}\right|^2.
\label{eq:sky_cal}
\end{equation}

The resulting calibration solutions are degenerate. They have one degenerate parameter (implicitly per frequency, time, and polarization) that corresponds to the overall phase of the gains. One can see this degeneracy by noting that the transformation $\gains \rightarrow \gains e^{i \phi}$ does not change the form of $\negloglikelihood(\gains)$ for any real value $\phi$. This phase can be interpreted as the absolute timing of incident radiation on the array and can be constrained by calibrating to a time-variable signal such as a pulsar \citep{Pen2009}.

All calibration frameworks described in this paper assume calibration to time-constant signals and therefore experience degeneracy in the overall phase of the gains. Typically this is constrained with a reference antenna. One then requires that $\operatorname{Arg}[\hat{\gains}_\text{ref}] = 0$ where $\hat{\gains}_\text{ref}$ is the maximum-likelihood gain of the reference antenna and $\operatorname{Arg}$ denotes the complex phase. However, this means that systematic errors in the phase of the reference antenna affect all antennas. More sophisticated calibration approaches mitigate calibration errors in the reference antenna to better constrain the overall phase of the gains \citep{Barry2019, Li2019}.

Traditional sky-based calibration assumes excellent knowledge of the sky and requires a highly complete sky model. It also assumes that the antenna responses are known up to a complex multiplicative factor. In other words, it assumes that the data can be fully modeled by the expression in Equation \ref{eq:vis_model_sky_cal}. This is clearly an inaccurate assumption. In reality, sky models are incomplete and inaccurate. These errors can propagate through the calibration process and degrade the calibration solutions \citep{Grobler2014,Barry2016,Patil2016, Ewall-Wice2017,Joseph2020}.

To account for sky model errors, we can replace the model visibilities with tunable calibration parameters $\fitparamsu$. Now $\fitparams = \{ \gains, \fitparamsu \}$ and
\begin{equation}
    \visfunc_{ab}=\gains_a \gains_b^* \fitparamsu_{ab}.  
\end{equation}
The likelihood function is now given by
\begin{equation}
    P(\gains, \fitparamsu|\data) \propto e^{-\frac{1}{2} \sum_{ab} \frac{1}{\sigma_{T,ab}^2} \left|v_{ab} - \gains_a \gains_b^* \fitparamsu_{ab}\right|^2} P(\fitparamsu),
\label{eq:sky_cal_partial_knowledge}
\end{equation}
where $P(\fitparamsu)$ is the prior on the tunable visibility parameters. Traditional sky-based calibration assumes that $\fitparamsu$ is known exactly. Equation \ref{eq:sky_cal_partial_knowledge} converges to Equation \ref{eq:sky_cal_likelihood} when 
\begin{equation}
    P(\fitparamsu) \propto \delta(\fitparamsu - \modelvals),
\label{eq:delta_func_prior}
\end{equation}
where $\delta$ denotes the Dirac delta function. One expects that a more accurate calibration framework would not require so stringent a prior on $\fitparamsu$. In \S\ref{s:general_framework} we explore the implications of relaxing this assumption.

\subsection{Redundant Calibration}
\label{s:red_cal}

Traditional redundant calibration \citep{Wieringa1992} uses the following choices:
\begin{enumerate}
    \item Like traditional sky-based calibration, traditional redundant calibration involves a tunable complex gain per antenna. However, it additionally assumes that the visibilities are not well-modeled and includes them as tunable sky model parameters. Redundant calibration works with highly regular arrays. In its traditional form, it assumes that baselines within a redundant set measure the same sky signal. The number of free visibility parameters is therefore equal to the number of redundant baseline sets. $\fitparams = \{\gains, \fitparamsu\}$ where $\fitparamsu$ correspond to the visibilities from redundant baseline sets. The number of parameters to fit in calibration is equal to the number of antennas plus the number of redundant baseline sets.
    \item It models the data as 
    \begin{equation}
        \visfunc_{ab}= \gains_a \gains_b^* \matrixa_{ab,j} \fitparamsu_j.
    \end{equation}
    Here $a$ and $b$ index antennas and $j$ indexes redundant baseline sets. $\matrixa$ is a matrix that maps $\fitparamsu$ to the full set of visibilities. Its elements are given by
    \begin{equation}
        \matrixa_{ab,j} = \begin{cases}
            1, & \text{if baseline $\{a,b\}$ belongs to set $j$} \\
            0, & \text{otherwise} 
        \end{cases}.
    \label{eq:A_trad_red_cal}
    \end{equation}    
    \item Like traditional sky-based calibration, traditional redundant calibration describes $P(\data|\fitparams)$ as an independent Gaussian probability distribution as given by Equation \ref{eq:independent_gaussian}.
    \item It uses a flat prior on the tunable calibration parameters such that $P(\fitparams) = P(\gains, \fitparamsu)$ is a constant.
\end{enumerate}

Under these assumptions, the likelihood function becomes
\begin{equation}
    P(\gains,\fitparamsu|\data) \propto e^{-\frac{1}{2} \sum_{ab} \sum_j \frac{1}{\thermalvar{}_{,ab}} \left|\data_{ab} - \gains_a \gains_b^* \matrixa_{ab,j} \fitparamsu_j \right|^2}.
\end{equation}
It follows that the negative log-likelihood is
\begin{equation}
    \negloglikelihood(\gains,\fitparamsu) = \sum_{ab} \sum_j \frac{1}{\thermalvar{}_{,ab}} \left|\data_{ab} - \gains_a \gains_b^* \matrixa_{ab,j} \fitparamsu_j \right|^2.
\label{eq:red_cal}
\end{equation}

The additional tunable calibration parameters in traditional redundant calibration, as compared to traditional sky-based calibration, introduce additional degeneracies. Not only are the calibration solutions degenerate in the overall phase of the gains but they are also degenerate in the overall gain amplitude and the gradient of the gains' complex phase across the array \citep{Liu2010}. These degeneracies mean that redundant calibration must be combined with `absolute calibration' to yield physical calibration solutions, where absolute calibration sets the values of the degenerate parameters. Minimizing Equation \ref{eq:red_cal} is sometimes called `relative calibration' to distinguish it from true redundant calibration, which must include both relative and absolute calibration steps \citep{Zheng2014, Byrne2019, Kern2019}.

The traditional redundant calibration framework assumes perfect redundancy of baselines within a redundant set. In order for different baselines within a redundant set to measure the same sky signal, antenna positions must lie on a perfect grid and each antennas' response must be identical up to a multiplicative gain. In practice, antenna position errors and response inhomogeneities degrade array redundancy \citep{Joseph2018,Li2018, Orosz2019}. In \S\ref{s:red_cal_imperfect_redundancy}-\S\ref{s:uv_cal} we propose new calibration frameworks that account for imperfect array redundancy.

Furthermore, traditional redundant calibration neglects covariance between redundant baseline sets. This is an appropriate assumption for some redundant arrays, such as the Donald C. Backer Precision Array for Probing the EoR (PAPER; \citealt{Parsons2010, Ali2015}) and the hexagonal sub-arrays in the MWA Phase II configuration \citep{Wayth2018}, but compact redundant arrays such as the Hydrogen EoR Array (HERA; \citealt{DeBoer2017}), the Hydrogen Intensity Real-Time Analysis Experiment (HIRAX; \citealt{Newburgh2016}), and the Canadian Hydrogen Observatory and Radio-transient Detector (CHORD; \citealt{Vanderlinde2019}) can have appreciable covariance between baselines with centers at small \textit{uv} separations. In \S\ref{s:red_cal_known_imperfect_redundancy}-\S\ref{s:uv_cal} we relax the assumption that baselines in different redundant baseline sets have zero covariance.

Redundant calibration can be an attractive alternative to sky-based calibration because it has reduced reliance on the sky model. One might think that redundant calibration can calibrate interferometric data without any prior knowledge of the sky --- after all, Equation \ref{eq:red_cal} doesn't require a sky model at all! However, this neglects the role of the sky model in absolute calibration. Although the promise of redundant calibration is to reduce the impact of sky model errors on calibration solutions, it is inaccurate to assume one can calibrate without \textit{any} prior knowledge of the sky. Furthermore, it is possible that introducing a sky model into the relative calibration step of redundant calibration as a prior on the fit visibility values could improve calibration performance. In \S\ref{s:general_framework} we explore calibration approaches that do just that.

\section{A General Framework for Direction-Independent Calibration}
\label{s:general_framework}

In this section we describe a novel and highly general calibration framework. This framework allows for extensions to traditional sky-based and redundant calibration that relax some of the non-physical assumptions of those approaches.

The calibration framework described in \S\ref{s:cal_recipe}-\S\ref{s:uv_cal} is inherently direction-independent. In \S\ref{s:direction_dependent} we discuss its extension to direction-dependent calibration. For a more in-depth discussion of direction-dependent calibration, see \citealt{Kazemi2011, Kazemi2013a, Patil2016, Weeren2016, Tasse2018, Albert2020}; or \citealt{Mertens2020}. Additionally, this calibration framework assumes per-frequency, per-time, and per-polarization calibration. \S\ref{s:freq_cal} extends the framework to frequency-dependent calibration and \S\ref{s:pol_cal} discusses fully polarized calibration techniques.

\subsection{General Framework}
\label{s:cal_recipe}

In its most general form, this calibration framework makes the following assumptions:
\begin{enumerate}
    \item It involves tunable complex gains per antenna (and implicitly per polarization, frequency, and time) $\gains$. The gains are not direction-dependent, meaning that this calibration framework is inherently direction-independent. In addition, this calibration framework involves tunable sky parameters $\fitparamsu$. The specific form of $\fitparamsu$ depends on the class of calibration used. $\fitparamsu$ can be interpreted as visibilities, pixels in the \textit{uv} plane, source flux densities, or something else altogether. In general, $\fitparamsu$ represents aspects of the sky model, possibly together with aspects of the instrument model, that are fit in calibration. The tunable calibration parameters are $\fitparams=\{ \gains, \fitparamsu\}$.

    \item We model the data as
    \begin{equation}
        \visfunc_{ab}=\gains_a \gains_b^* \sum_j \matrixa_{ab, j} \fitparamsu_j. 
    \end{equation}
    Here $\matrixa$ is a matrix that maps the sky parameters $\fitparamsu$ to model visibilities. In more compact matrix multiplication notation, we can write 
    \begin{equation}
        \visfunc = \gainsmat(\gains) \matrixa \fitparamsu,
    \label{eq:vis_model}
    \end{equation}
    where we have defined $\gainsmat(\gains)$ as a diagonal matrix with elements $\gains_a \gains_b^*$. In the case that $\fitparamsu$ represents visibilities, $\matrixa$ is simply the identity matrix: $\matrixa = \mathbb{1}$. 

    \item We describe $P(\data|\fitparams)$ as an independent Gaussian probability distribution of the form of Equation \ref{eq:independent_gaussian}.

    \item We use a Gaussian prior on $\fitparamsu$ such that
    \begin{equation}
        P(\fitparamsu) \propto e^{-\frac{1}{2} (\fitparamsu - \modelvals)^\dag \modelcov^{-1} (\fitparamsu - \modelvals)}
    \label{eq:u_prior}
    \end{equation}
    where $\modelvals = \langle \fitparamsu \rangle$. $\modelcov = \operatorname{cov}[\fitparamsu, \fitparamsu^\dag]$ is an invertible matrix that encodes the covariances between elements of $\fitparamsu$. Furthermore, this calibration framework assumes that $\gains$ and $\fitparamsu$ are independent and uses a flat prior on $\gains$:
    \begin{equation}
        P(\fitparams) = P(\gains, \fitparamsu) = P(\gains)P(\fitparamsu) \propto P(\fitparamsu).
    \end{equation}
\end{enumerate}

Under this general framework, the likelihood function is given by
\begin{equation}
\begin{split}
    P(\gains, \fitparamsu|\data) \propto & e^{-\frac{1}{2}[\data - \gainsmat(\gains) \matrixa \fitparamsu]^\dag \thermalcov^{-1} [\data - \gainsmat(\gains) \matrixa \fitparamsu]} \\
    & \times e^{-\frac{1}{2}(\fitparamsu - \modelvals)^\dag \modelcov^{-1} (\fitparamsu - \modelvals)}
\end{split}
\end{equation}
and $\negloglikelihood(\gains,\fitparamsu)$ is given by
\begin{equation}
\begin{split}
    \negloglikelihood(\gains,\fitparamsu) = &[\data - \gainsmat(\gains) \matrixa \fitparamsu]^\dag \thermalcov^{-1} [\data - \gainsmat(\gains) \matrixa \fitparamsu] \\
    &+ (\fitparamsu - \modelvals)^\dag \modelcov^{-1} (\fitparamsu - \modelvals).
\end{split}
\label{eq:red_cal_most_general}
\end{equation}
$\thermalcov$ is diagonal because the thermal noise on the visibilities is assumed to be independent.

The calibration framework described here is highly abstracted. It represents a generalized treatment can be applied to practical calibration problems. In \S\ref{s:sky_cal_ex}-\S\ref{s:uv_cal} we delineate examples of calibration approaches that emerge from this framework. These are just some of the possible avenues for future exploration.

\subsection{Sky-Based Calibration with Partial Sky Model Knowledge}
\label{s:sky_cal_ex}

Traditional sky-based calibration, presented in \S\ref{s:sky_cal}, assumes that the model of the visibilities represents the true signal up to a multiplicative complex gain. Of course this is not true. Model visibilities are susceptible to errors from incomplete knowledge of the sky and instrument. We can make sky-based calibration more resilient to sky model errors by including tunable visibility parameters $\fitparamsu$ in calibration.

A dominant source of model visibility error stems from faint missing sources in the sky model catalog. We assume many sources exist with intensities below the sensitivity limit of the catalog. Each of these sources produces a fringe pattern in \textit{uv} space. The signals from different sources add incoherently. In the limit of many sources randomly distributed on the sky, with flux densities randomly drawn from a power-law distribution, the central limit theorem dictates that the combined signal approaches Gaussian random noise in the \textit{uv} plane \citep{Tegmark1998}. This suggests that the true visibilities are Gaussian-distributed around their model values and motivates using a Gaussian prior $P(\fitparamsu)$ given by Equation \ref{eq:u_prior}. This assumption breaks down if model visibility errors are instead dominated by a small number of bright sources that are either missing or mis-modeled in the sky model catalog. \S\ref{s:variable_source_intensities} presents a calibration framework that assumes correctly modeled source positions but uncertain source intensities. 

If the baseline measurements sample discrete regions of \textit{uv} space we further expect that the visibilities are independent. This means that $\modelcov$ is diagonal. Compact or redundant arrays can have significant baseline overlap in \textit{uv} space and the assumption of independent visibilities does not hold. In \S\ref{s:red_cal_with_sky_model}-\S\ref{s:uv_cal} we explore additional calibration frameworks that account for visibility covariance.

Under these assumptions, the prior on the tunable visibility parameters $\fitparamsu$ takes the form
\begin{equation}
    P(\fitparamsu) \propto e^{-\frac{1}{2} \sum_j \frac{1}{\modelvar{}_{,j}} |\fitparamsu_j - \modelvals_j|^2},
\end{equation}
where $\modelvar{}_{,j}$ are the diagonal elements of $\modelcov$. We therefore get $\negloglikelihood(\gains,\fitparamsu)$ of the form
\begin{equation}
\begin{split}
    \negloglikelihood(\gains,\fitparamsu) = &\sum_{ab} \frac{1}{\thermalvar{}_{,ab}} \left|\data_{ab} - \gains_a \gains_b^* \fitparamsu_{ab}\right|^2 \\
    &+ \sum_j \frac{1}{\modelvar{}_{,j}} |\fitparamsu_j - \modelvals_j|^2
\end{split}
\label{eq:sky_cal_ex}
\end{equation}
and fit for calibration parameters $\fitparams = \{ \gains, \fitparamsu \}$.
Note that this converges to traditional sky calibration (Equation \ref{eq:sky_cal}) in the limit that $\modelvar{}_{,j} \rightarrow 0$.

The addition of new tunable calibration parameters $\fitparamsu$ expands the number of tunable parameters as compared to traditional sky-based calibration (\S\ref{s:sky_cal}). Traditional sky-based calibration fits one complex parameter per antenna for a total of $N_\text{ant}$ complex parameters, where $N_\text{ant}$ refers to the number of antennas. Minimizing Equation \ref{eq:sky_cal_ex} involves fitting an additional number of parameters equal to the number of independent baseline measurements. As autocorrelation visibilities are typically omitted from calibration, this corresponds to $(N_\text{ant}^2 - N_\text{ant})/2$ additional complex parameters. Calibration runtimes are highly dependent on the number of tunable parameters with the precise relationship depending on the implementation details.

We note that while this calibration approach involves many tunable parameters, overfitting is not a concern because the tunable visibility parameters $\fitparamsu$ are constrained by their prior $P(\fitparamsu)$. While their maximum-likelihood values may fit some degree of thermal noise, this noise does not impact the gains $\gains$ to a greater extent than it would in a traditional sky-based calibration approach.

One common sky-based calibration technique is to calibrate only on selected baselines where the sky model is trusted. For example, this could involve calibrating only on baselines longer than 50 wavelengths in order to eliminate the short baselines that are poorly modeled by a point source catalog. We can replicate this technique from the formalism described in Equation \ref{eq:sky_cal_ex} by taking the limits
\begin{equation}
    \modelvar{}_{,j} \rightarrow \begin{cases}
    0 & \text{if baseline $j$ is included in calibration} \\
    \infty & \text{if baseline $j$ is excluded from calibration}
    \end{cases}.
\end{equation}
We now find that $\fitparamsu_j \rightarrow \modelvals_j$ for those baselines that are included in calibration: the fitted visibilities for those baselines are constrained to match the model visibilities. On the other hand, $\fitparamsu_j$ for baselines that are excluded from calibration are completely unconstrained. They will therefore take on values such that $\gains_a \gains_b^* \fitparamsu_{ab} = \data_{ab}$ for any values of the gains $\gains_a$ and $\gains_b$.

While this calibration framework can replicate binary baseline selection it need not completely include or exclude baselines. $\modelvar$ quantifies the uncertainty on the model visibilities and can take any values. For example, it can be adjusted to be a function of baseline length to represent different levels of model confidence on different angular scales \citep{Ewall-Wice2017}. Rather than completely eliminating short baselines from calibration, one could instead selectively downweight them, gradually increasing the value of $\modelvar{}_{,j}$ on subsequently shorter and shorter baselines. $\modelvar$ could also be calculated empirically by measuring the agreement of the data and sky model as a function of baseline length.

\subsection{Sky-Based Calibration with Uncertain Source Intensitites}
\label{s:variable_source_intensities}

Some calibration approaches in the literature consider the case of a point source sky model in which source intensities are not well-constrained \citep{Mitchell2008, Sievers2017}. In this case, it could be advantageous to redefine $\fitparamsu$ to be the source intensities rather than visibilities. Rather than assuming that sky model errors are dominated by many unmodeled sources with random positions, as in \S\ref{s:sky_cal_ex}, this framework assumes that the source catalog is highly complete with well-modeled source positions. It could work well when the dominant sky model errors stem from errors in the modeled flux density of known sources. Under these assumptions, we can define $\negloglikelihood(\gains,\fitparamsu)$ to take the form
\begin{equation}
\begin{split}
    \negloglikelihood(\gains,\fitparamsu) = &[\data - \gainsmat(\gains) \matrixa \fitparamsu]^\dag \thermalcov^{-1} [\data - \gainsmat(\gains) \matrixa \fitparamsu] \\
    &+ \sum_j \frac{1}{\modelvar{}_{,j}} |\fitparamsu_j - \modelvals_j|^2,
\end{split}
\end{equation}
where $\modelvals_j = \langle \fitparamsu_j \rangle$ is the expected flux density of source $j$ and $\modelvar{}_{,j}$ quantifies the uncertainty of that intensity. The matrix $\matrixa$ models the instrument response. The element $\matrixa_{jk}$ is equal to the contribution to visibility $j$ from a source of unity intensity at the position of source $k$.

Precision calibration for 21cm cosmology often uses tens of thousands of calibrator sources in the model. Allowing the intensities of all sources to vary is computationally infeasible. Instead, a more realistic approach would be to take $\modelvar{}_{,j} \rightarrow 0$, such that $\fitparamsu_j \rightarrow \modelvals_j$, for the majority of sources $j$. Then $\modelvar{}_{,j} > 0$ only for a select subset of troublesome sources that one expects are inaccurately modeled.

This calibration approach is similar to direction-dependent calibration (\S\ref{s:direction_dependent}) in that it can account for mis-modeled apparent source intensities that vary across the sky. However, we classify this approach as direction-independent because the fitted gains --- the calibration parameters that are actually applied to the data --- are not direction-dependent. Importantly, this calibration framework does not fit a per-antenna intensity variation across the sky.

\subsection{Redundant Calibration With a Sky Model and Calibration of Hybrid (Redundant and Non-Redundant) Arrays}
\label{s:red_cal_with_sky_model}

As explained in \S\ref{s:red_cal}, traditional redundant calibration does not require a sky model or an explicit instrument model in the relative calibration step. However, this means that it yields degenerate solutions. These degeneracies must be broken through an absolute calibration step that fits the degenerate parameters to model visibilities.

An extension to traditional redundant calibration incorporates model visibilities in $\negloglikelihood(\gains,\fitparamsu)$. This eliminates the need for a separate absolute calibration step and instead performs relative and absolute calibration simultaneously. It can also constrain the tunable visibility parameters $\fitparamsu$ to approximate a known sky model.

As in \S\ref{s:red_cal} we define $\fitparamsu$ to be the independent visibilities with length equal to the number of redundant baseline sets. We assume that different redundant baseline sets are independent and that, as in \S\ref{s:sky_cal_ex}, the errors on the model \textit{uv} plane are Gaussian distributed. Under these assumptions, we expand the traditional redundant calibration framework given by Equation \ref{eq:red_cal} by including a Bayesian prior on $\fitparamsu$:
\begin{equation}
\begin{split}
    \negloglikelihood(\gains,\fitparamsu) = &[\data - \gainsmat(\gains) \matrixa \fitparamsu]^\dag \thermalcov^{-1} [\data - \gainsmat(\gains) \matrixa \fitparamsu] \\
    &+ \sum_j \frac{1}{\modelvar{}_{,j}} |\fitparamsu_j - \modelvals_j|^2.
\end{split}
\label{eq:red_cal_plus_sky}
\end{equation}
$\matrixa$ maps the independent visibilites to the full set of visibilities and is given by Equation \ref{eq:A_trad_red_cal}.

At this point it is useful to note that Equation \ref{eq:red_cal_plus_sky} can be equivalently derived by considering the full basis of visibilities. Instead of defining $\fitparamsu$ as the visibilities from redundant baseline sets, we instead consider $\fitparamsu_\text{orig}$, with length equal to the total number of visibilities. The associated matrix $\matrixa_\text{orig} = \mathbb{1}$. The visibility covariance matrix $\modelcov{}_\text{,orig} = \text{cov}[\fitparamsu_\text{orig}, \fitparamsu_\text{orig}^\dag]$. If we assume perfect redundancy, such that baselines within a redundant set measure the same signal, then $\text{cov}[\fitparamsu_{\text{orig}, j}, \fitparamsu_{\text{orig}, k}^*] = \text{var}[\fitparamsu_{\text{orig}, j}]$ when baselines $j$ and $k$ belong to the same redundant baseline set. If we further assume independence between visibilities from different redundant baseline sets then $\text{cov}[\fitparamsu_{\text{orig}, j}, \fitparamsu_{\text{orig}, k}^*] = 0$ when baselines $j$ and $k$ belong to different redundant baseline sets. We thereby derive a block-diagonal matrix $\modelcov{}_\text{,orig}$ where the number of blocks corresponds to the number of redundant baseline sets.

However, we would encounter a problem deriving $\negloglikelihood(\gains,\fitparamsu_
\text{orig})$ by plugging these quantities into Equation \ref{eq:red_cal_most_general}. $\modelcov{}_\text{,orig}$ is singular: it has rank equal to the number of redundant baseline sets. This belies that $\fitparamsu_\text{orig}$ is not a proper basis for calibration. It must be remapped into a new basis, which we derive by applying Singular Value Decomposition (SVD) to $\modelcov{}_\text{,orig}$ (see Appendix \ref{app:perfect_redundancy}). This generates a new calibration basis $\fitparamsu$ where $\fitparamsu_
\text{orig} = \matrixa \fitparamsu$ and $\matrixa$ is given by Equation \ref{eq:A_trad_red_cal}. We thereby recover the calibration framework given in Equation \ref{eq:red_cal_plus_sky}.

The calibration framework described in this section unifies the `absolute' and `relative' steps of redundant calibration. Like traditional sky-based calibration (\S\ref{s:sky_cal}), it is degenerate only in the overall phase of the gains. Yet, like traditional redundant calibration (\S\ref{s:red_cal}), it incorporates information from baseline redundancy.

This framework involves $N_\text{ant} + N_\text{red}$ tunable calibration parameters, where $N_\text{red}$ refers to the number of redundant baseline sets. This is equivalent to the number of tunable parameters used in traditional redundant calibration.

The framework also has interesting implications for `hybrid' arrays, i.e.\ arrays with some redundant and some non-redundant elements. Phase II of the MWA, for example, contains two hexagonal sub-arrays that support redundant calibration. These redundant sub-arrays are contained within a larger pseudo-random array. Thus far, calibrating a hybrid array has required separate calibration operations for the redundant and non-redundant elements \citep{Li2018,Li2019}. In order to exploit the redundancy of the sub-arrays one had to relatively calibrate those antennas with traditional redundant calibration techniques, therefore using baseline measurements only from within the sub-arrays. The calibration framework described in this section allows for calibration of the redundant sub-array antennas using all baseline measurements. This incorporates information from all baselines involving that antenna and can increase the calibration signal-to-noise.

To explore the scope of this calibration approach we describe two opposing limits of the calibration model. The first limit takes $\modelvar \ll \thermalvar$. In this limit we assume that the model visibilities $\modelvals$ are well-known and require that $\fitparamsu \approx \modelvals$. Extending this limit such that $\modelvar \rightarrow 0$ enforces that $\fitparamsu \rightarrow \modelvals$. Equation \ref{eq:red_cal_plus_sky} is thereby equal Equation \ref{eq:sky_cal}, and we recover traditional sky-based calibration.

In the opposing limit $\modelvar \gg \thermalvar$, where $\thermalvar$ are elements of the diagonal matrix $\thermalcov$. The model visibilities $\modelvals$ will then have a negligible contribution to the relative calibration parameters, i.e.\ the calibration parameters constrained by minimizing Equation \ref{eq:red_cal}. Instead, the model visibilities will fit the absolute calibration parameters only. This limit corresponds to a unified redundant calibration framework where absolute and relative calibration are encompassed in a single minimization problem. We note that extending this limit such that $\modelvar \rightarrow \infty$ eliminates the second term in Equation \ref{eq:red_cal_plus_sky}. We thereby recover Equation \ref{eq:red_cal} and traditional redundant calibration.

In practice, neither of these limits are physically-motivated. $\modelvals$ cannot perfectly model the visibilities; on the other hand, it \textit{can} give some information about the relative calibration solutions. The strength of this calibration approach is that it can inhabit the middle ground of \textit{some} confidence in the model visibilities. As in \S\ref{s:sky_cal_ex}, we could set $\modelvar$ empirically by measuring discrepancies between the data and model. In \S\ref{s:sim_results} we present simulation results that explore the implications of working in an intermediate regime between traditional redundant and sky-based calibration.

\subsection{Redundant Calibration Accounting for Unmodeled Imperfect Redundancy}
\label{s:red_cal_imperfect_redundancy}

As explained in \S\ref{s:red_cal_with_sky_model}, we can describe traditional redundant calibration with a block-diagonal $\modelcov$ when the basis of sky parameters $\fitparamsu$ corresponds to the full set of visibilities (denoted $\modelcov{}_\text{,orig}$ in \S\ref{s:red_cal_with_sky_model}). Traditional redundant calibration assumes $\operatorname{cov}[\fitparamsu_j, \fitparamsu_k^*] = \operatorname{var}[\fitparamsu_j]$ for visibilities $\fitparamsu_j$ and $\fitparamsu_k$ if baselines $j$ and $k$ belong to the same redundant set.

However, it is reasonable to assume that small antenna position errors and beam response inhomogeneities make $\operatorname{cov}[\fitparamsu_j, \fitparamsu_k^*] < \operatorname{var}[\fitparamsu_j]$ when $j \neq k$. We can account for this by applying a suppression term to the off-diagonal elements of $\modelcov$. This suppression factor could be calculated empirically by measuring the covariance between visibilities from redundant baselines as the array measures different fields on the sky. It could also be baseline- or antenna-dependent. For example, if one antenna is known to have a particularly irregular beam response, the baseline covariance terms for baselines that include that antenna could be preferentially suppressed.

In general, suppressing the off-diagonal elements of a block-diagonal $\modelcov$ renders it invertible (in theory at least; in practice inverting $\modelcov$ could be computationally prohibitive). We may no longer need to recast $\fitparamsu$ to a reduced basis such as the set of independent visibilities or the Singular Value Decomposition (SVD) basis described in \S\ref{app:perfect_redundancy}. However, under this calibration approach the tunable visibility parameters $\fitparamsu$ will be highly correlated. The maximum likelihood values $\hat{\fitparamsu}$ associated with redundant baselines will, by construction, have similar values. For this reason, we cannot use agreement of $\hat{\fitparamsu}$ for redundant baselines as evidence of the baselines' true degree of redundancy.

That said, considering the SVD basis can be enlightening for understanding the effect of suppressing the off-diagonal elements of $\modelcov$ by a small amount. The additional calibration degrees-of-freedom introduced correspond to differences in $\fitparamsu$ between visibility measurements within a redundant baseline set. The variance on these differences is small. Provided the model visibilities $\modelvals$ are the same for all visibilities from a redundant set, the differences between visibility measurements in a redundant set are constrained to be near-zero.

Suppressing all off-diagonal elements of $\modelcov$ increases the number of tunable calibration parameters to $(N_\text{ant}^2 + N_\text{ant})/2$, as compared to $N_\text{ant} + N_\text{red}$ parameters for traditional redundant calibration. For large arrays this could substantially increase in calibration runtimes.

\subsection{Redundant Calibration with Modeled Imperfect Redundancy}
\label{s:red_cal_known_imperfect_redundancy}

\S\ref{s:red_cal_imperfect_redundancy} describes a calibration framework that accounts for unmodeled antenna position errors and beam response inhomogeneities. However, often antenna positions and beam responses can be measured to greater accuracy than they can be controlled. For example, an antenna in a redundant array may have a known position offset from its ideal position. In this case, the covariance matrix $\modelcov$ can be calculated from the modeled \textit{uv} responses of the antennas.

We represent the model of the \textit{uv} response of baseline $j$ as $\beamvec_j(\uvcoord)$, where $\uvcoord$ is the \textit{uv} position vector. These baseline response models can be developed from beam simulators or direct measurements and are readily available from instrument simulators such as Fast Holographic Deconvolution\footnote{\texttt{https://github.com/EoRImaging/FHD}} (FHD; \citealt{Sullivan2012}), pyuvsim\footnote{\texttt{https://github.com/RadioAstronomySoftwareGroup/pyuvsim}} \citep{Lanman2019}, OSKAR\footnote{\texttt{https://github.com/OxfordSKA/OSKAR}} \citep{Mort2010}, Precision Radio Interferometry Simulator\footnote{\texttt{https://github.com/nithyanandan/PRISim}} (PRISim), or Common Astronomy Software Applications (CASA; see \citealt{Jagannathan2017} for an example of using CASA with a fully-polarized primary beam model). The visibility $\fitparamsu_j$ relates to the true \text{uv} plane $\boldsymbol{S}(\uvcoord)$ via
\begin{equation}
    \fitparamsu_j = \int_{-\infty}^\infty \beamvec_j(\uvcoord) \boldsymbol{S}(\uvcoord) d^2\uvcoord.
\end{equation}

If we assume no uncertainty on the \textit{uv} response models, the elements of the covariance matrix $\modelcov$ are given by
\begin{equation}
\begin{split}
    &\modelcov{}_{,jk} = \operatorname{cov}[\fitparamsu_j, \fitparamsu_k^*] = \\
    &\int_{-\infty}^\infty \int_{-\infty}^\infty \beamvec_j(\uvcoord) \beamvec_k^*(\uvcoord') \operatorname{cov}[\boldsymbol{S}(\uvcoord), \boldsymbol{S}^*(\uvcoord')] d^2\uvcoord d^2\uvcoord'.
\end{split}
\end{equation}
As in \S\ref{s:sky_cal_ex}, we assume that sky model errors are dominated by many faint sources missing from the model catalog. We expect the resulting error in the \textit{uv} plane to be Gaussian-distributed and fully described by a variance $\sigma_\text{M}^2$ at each point. We further assume that points in the \textit{uv} plane are independent:
\begin{equation}
    \operatorname{cov}[\boldsymbol{S}(\uvcoord), \boldsymbol{S}^*(\uvcoord')] = \begin{cases}
    0 & \text{when $\uvcoord \neq \uvcoord'$} \\
    \sigma_\text{M}^2 & \text{when $\uvcoord = \uvcoord'$}
    \end{cases}.
\end{equation}
Under these assumptions, we get that
\begin{equation}
    \modelcov{}_{,jk} = \sigma_\text{M}^2 \int_{-\infty}^\infty \beamvec_j(\uvcoord) \beamvec_k^*(\uvcoord) d^2\uvcoord.
\label{eq:cov_calc_integral}
\end{equation}

This method of calculating elements of the covariance matrix allows us to accurately represent all amounts of modeled baseline covariance in the calibration model. Redundant baselines have highly overlapping \textit{uv} responses. In a closely packed array, baselines that are not constructed to be redundant may nonetheless overlap somewhat in \textit{uv} coverage; this overlap introduces nonzero covariance in their measurements. Traditional redundant calibration (\S\ref{s:red_cal}) assumes that $\beamvec_j(\uvcoord) = \beamvec_k(\uvcoord)$ when baselines $j$ and $k$ belong to the same redundant baseline set. It also assumes that when $j$ and $k$ do not belong to the same redundant baseline set their \textit{uv} coverage does not overlap at all: $\int_{-\infty}^\infty \beamvec_j(\uvcoord) \beamvec_k^*(\uvcoord) d^2\uvcoord = 0$. If we instead calculate a covariance matrix from Equation \ref{eq:cov_calc_integral} we can relax those assumptions and build a more physically-motivated calibration model that accounts for array non-redundancies and can incorporate covariances from baseline overlap in closely packed arrays such as HERA \citep{DeBoer2017} and HIRAX \citep{Newburgh2016}.

In \S\ref{s:sky_cal_ex} we noted that a sky model may be more trusted for some baselines than others. The assumption in Equation \ref{eq:cov_calc_integral} of uniform variance across the \textit{uv} plane therefore must be relaxed. If we instead assume that the variance of the \textit{uv} plane is approximately constant at scales equal to the size of a baseline response, we can rewrite Equation \ref{eq:cov_calc_integral} as
\begin{equation}
    \modelcov{}_{,jk} = \boldsymbol{\sigma}_{\text{M},j} \boldsymbol{\sigma}_{\text{M},k} \int_{-\infty}^\infty \beamvec_j(\uvcoord) \beamvec_k^*(\uvcoord) d^2\uvcoord.
\label{eq:cov_calc_integral_ext}
\end{equation}
Here $\modelvar{}_{,j}$ is simply the variance of $\fitparamsu_j$; $\boldsymbol{\sigma}_{\text{M},j}$ is the standard deviation. As noted in \S\ref{s:sky_cal_ex}, one could determine values of $\boldsymbol{\sigma}_{\text{M}}$ empirically by comparing the data to the model. Poorly-modeled power spectrum modes could be downweighted in calibration by increasing $\boldsymbol{\sigma}_{\text{M}, j}$ for the associated baseline lengths. A baseline $j$ could be excluded from calibration altogether by taking $\boldsymbol{\sigma}_{\text{M},j} \rightarrow \infty$.

We can combine the calibration approach described in this section with that of \S\ref{s:red_cal_imperfect_redundancy}. One could expect that further non-redundancies exist beyond what is encoded in the baseline response model $\beamvec_j(\uvcoord)$. In that case, one could represent non-modeled antenna position or beam response errors by suppressing the off-diagonal elements of $\modelcov$. As in \S\ref{s:red_cal_imperfect_redundancy}, this suppression factor could be estimated from empirical measurements of the covariance of visibilities. Calculating visibility covariances across observations of different parts of the sky could help validate or adjust a covariance matrix constructed from Equation \ref{eq:cov_calc_integral}.

As in \S\ref{s:red_cal_imperfect_redundancy}, under the calibration framework presented in this section the number of tunable calibration parameters lies between $N_\text{ant}+N_\text{red}$ and $(N_\text{ant}^2+N_\text{ant})/2$, with the exact number depending on whether any baselines are modeled as fully redundant. The number of tunable parameters can be reduced by eliminating the most constrained calibration degrees-of-freedom. This is analogous to deciding that the most redundant baselines can be approximated as fully redundant. Formally, this is accomplished by performing Principal Value Decomposition (PVD) on the covariance matrix $\modelcov$. We calculate the SVD basis according to Appendix \ref{app:perfect_redundancy}. Of the resulting parameters, we can choose not to calibrate the parameters with the smallest variances.

\subsection{Covariant Calibration of Compact Non-Redundant Arrays}
\label{s:compact_array_cal}

The calibration formalism presented in \S\ref{s:red_cal_known_imperfect_redundancy} is fully generalizable to a compact non-redundant array. The framework allows us to incorporate information from baseline covariance to constrain measurements across a non-redundant array.

For example, Phase I of the MWA \citep{Tingay2013} has no redundant baselines. As an imaging array, it was built to be pseudo-random with maximal \textit{uv} coverage, in a sense making it as non-redundant as possible \citep{Beardsley2012}. At the same time, it has a highly compact core and is \textit{uv} complete out to about 50 wavelengths. This high \textit{uv} coverage means that baselines measurements are not independent. Instead, many baselines are highly covariant with one another.

We can incorporate this covariance into the calibration framework by constructing $\modelcov$ from Equation \ref{eq:cov_calc_integral}. Fit visibility values $\fitparamsu$ would then be constrained not only by their modeled values $\modelvals$ but also by the visiblities from overlapping baselines in the \textit{uv} plane.

In this way the ideas behind redundant calibration can be extended to non-redundant compact arrays. Baselines need not be highly redundant to exhibit appreciable covariance, and this covariance can be used to constrain calibration solutions. These calibration techniques have applications for many classes of interferometers, not just highly redundant arrays.

\subsection{uv-Space Calibration}
\label{s:uv_cal}

Traditional redundant calibration interprets the tunable sky parameters $\fitparamsu$ to be visibilities (denoted $\fitparamsu_\text{orig}$ in this section), however one could instead reinterpret $\fitparamsu$ as pixels of the \textit{uv} plane. This is a natural parameter space for describing the model and means that $\modelcov$ does not depend on the instrument response. Instead, $\matrixa$ encodes degridding by mapping the pixels of the \textit{uv} plane to the visibilitites.

In \S\ref{s:red_cal_known_imperfect_redundancy} we describe the instrument response model as a continuous function in the \textit{uv} plane, but in practice the \textit{uv} plane is discretized. We can define a covariance matrix $\modelcov{}_{,\text{orig}}$ by rewriting Equation \ref{eq:cov_calc_integral} for a discrete \textit{uv} plane:
\begin{equation}
    \modelcov{}_{,\text{orig}} = \sigma_\text{M}^2 \beammat \beammat^\dag
\end{equation}
where $\beammat$ is a rectangular matrix that maps \textit{uv} pixels to visibilities. $\beammat$ has a number of columns equal to the number of pixels in the \textit{uv} plane and a number of rows equal to the number of visibilities. The product $\beammat \beammat^\dag$ is equivalent to the holographic mapping function described in \citealt{Sullivan2012}.

We define a new set of sky parameters $\fitparamsu$ that correspond to pixels in the \textit{uv} plane. $\beammat$ defines the mapping between $\fitparamsu$ and $\fitparamsu_\text{orig}$:
\begin{equation}
    \fitparamsu_\text{orig} = \beammat \fitparamsu.
\end{equation}
We now get that 
\begin{equation}
\begin{split}
    \modelcov{}_{,\text{orig}} &= \langle \fitparamsu_\text{orig} \fitparamsu_\text{orig}^\dag \rangle - \langle \fitparamsu_\text{orig} \rangle \langle \fitparamsu_\text{orig}^\dag \rangle \\
    &= \beammat ( \langle \fitparamsu \fitparamsu^\dag \rangle - \langle \fitparamsu \rangle \langle \fitparamsu^\dag \rangle ) \beammat^\dag \\
    &= \sigma^2_\text{M} \beammat \beammat^\dag,
\end{split}
\end{equation}
so the new covariance matrix is
\begin{equation}
    \modelcov = \langle \fitparamsu \fitparamsu^\dag \rangle - \langle \fitparamsu \rangle \langle \fitparamsu^\dag \rangle = \sigma^2_\text{M} \mathbb{1}.
\end{equation}
The fact that this covariance matrix is diagonal highlights the assumption in \S\ref{s:red_cal_known_imperfect_redundancy} of independent \textit{uv} pixels.

Plugging this new parameterization of $\fitparamsu$ into $\negloglikelihood(\gains,\fitparamsu)$ gives
\begin{equation}
\begin{split}
    \negloglikelihood(\gains,\fitparamsu) = &[\data - \gainsmat(\gains) \beammat \fitparamsu]^\dag \thermalcov^{-1} [\data - \gainsmat(\gains) \beammat \fitparamsu] \\
    &+ \frac{1}{\sigma^2_\text{M}} \sum_j |\fitparamsu_j - \modelvals_j|^2,
\end{split}
\label{eq:uv_cal}
\end{equation}
where $\modelvals$ is the model of the \textit{uv} plane and $j$ indexes over \textit{uv} plane pixels. We can also relax the assumption that $\modelvar{}_{,j} = \sigma_\text{M}^2$ for all \textit{uv} pixels $j$ and rewrite Equation \ref{eq:uv_cal} as
\begin{equation}
\begin{split}
    \negloglikelihood(\gains,\fitparamsu) = &[\data - \gainsmat(\gains) \beammat \fitparamsu]^\dag \thermalcov^{-1} [\data - \gainsmat(\gains) \beammat \fitparamsu] \\
    &+ \sum_j \frac{1}{\modelvar{}_{,j}} |\fitparamsu_j - \modelvals_j|^2.
\end{split}
\end{equation}
Now $\modelvar$ can represent variations in the quality of the sky model across the \textit{uv} plane. A point source sky model that omits diffuse emission could have $\modelvar$ increase for shorter baselines. Regions of \textit{uv} space could be excluded from calibration altogether by taking the limit $\modelvar{}_{,j} \rightarrow \infty$ for those pixels $j$. One could calculate $\modelvar$ empirically by comparing the model of the \textit{uv} plane to measurements.

We call this new calibration framework `\textit{uv}-space calibration.' It offers an alternative formulation of the calibration framework described in \S\ref{s:red_cal_known_imperfect_redundancy}. Parameterizing $\fitparamsu$ as \textit{uv} pixels rather than visibilities highlights the implicit assumption of statistical independence of the pixels by making $\modelcov$ diagonal. Here $\modelvals$ depends on the sky model only and is calculated without an instrument simulator. The instrument model is moved from the second to the first term of $\negloglikelihood(\gains,\fitparamsu)$, where $\beammat$ encodes the baseline responses to the \textit{uv} plane. Baseline covariances need not be calculated explicitly. Instead, this calibration framework implicitly constrains visiblities from fully- and partially-redundant baselines from their \textit{uv} plane overlap.

\textit{uv}-space calibration is a natural framework for unified calibration that models all baseline covariances. As with the calibration framework described in \S\ref{s:red_cal_known_imperfect_redundancy}, it can account for small non-redundancies of a redundant array due to antenna position offsets and beam inhomogeneities. Like in \S\ref{s:compact_array_cal}, it models covariances of non-redundant baselines stemming from baseline overlaps in the \textit{uv} plane. Note that \textit{uv}-space calibration incorporates modeled covariances only; it cannot include unmodeled imperfect redundancy as in \S\ref{s:red_cal_imperfect_redundancy}. Finally, like the calibration framework described in \S\ref{s:sky_cal_ex}, \textit{uv}-space calibration can incorporate variable sky model uncertainties as a function of \textit{uv} position to represent sky model incompleteness and preferentially fit calibration solutions to well-modeled \textit{uv} modes.

\subsection{Direction-Dependent Calibration}
\label{s:direction_dependent}

Direction-independent calibration assumes that the shape of the antenna responses across the sky are well-modeled \textit{a priori}. In practice, unmodeled direction-dependent effects can degrade interferometric performance. Direction-dependent calibration is a more flexible alternative that can capture uncertainties in the spatial structure of the antenna responses.

Direction-dependent calibration has been explored extensively in the literature. In this section we show that generalized direction-dependent calibration can be formally described using the statistical framework presented in this paper. For a more thorough discussion of direction-dependent calibration see \citealt{Kazemi2011, Kazemi2013a, Patil2016, Weeren2016, Tasse2018, Albert2020}; and \citealt{Mertens2020}.

A direction-dependent extension to traditional sky-based calibration (\S\ref{s:sky_cal}) uses the following assumptions:
\begin{enumerate}
    \item Like all calibration frameworks presented in this paper, direction-dependent calibration uses tunable complex gains per antenna --- but now, those gains can take different values at different positions on the sky. The tunable calibration parameters are given by $\fitparams = \{\gains_1, \gains_2, \dots, \gains_N \}$ for $N$ discrete sky sections, or facets. Here $\gains_n$ is the set of all gains for sky location $n$ and has length equal to the number of antennas.
    \item With direction-dependent calibration the visibility model $\visfunc(\fitparams)$ no longer takes the simple form of Equation \ref{eq:vis_model_sky_cal}. As in \S\ref{s:general_framework}, we can represent the mapping from sky parameters to visibilities with a matrix $\matrixa$, but $\matrixa$ is now a function of the direction-dependent gains. We can write the visibility model as
    \begin{equation}
        \visfunc(\gains_1, \gains_2, \dots, \gains_N) = \matrixa(\gains_1, \gains_2, \dots, \gains_N)\modelvals.
    \end{equation}
    Here $\modelvals$ represents the sky model. It could be parameterized as point source intensities or the intensities of pixels on the sky. Unlike in traditional sky-based calibration, $\modelvals$ cannot be parameterized as the estimate of the visibilities because the visibilities cannot be modeled \textit{a priori}. Simulated visibilities will depend on the direction-dependent gains.
    \item As in \S\ref{s:sky_cal}, one can describe $P(\data|\fitparams)$ as an independent Gaussian probability distribution (Equation \ref{eq:independent_gaussian}). This framework can be extended to other distributions with no loss of generality.
    \item We could use a flat prior for the direction-dependent gains. Alternatively, a common prior in direction-dependent calibration requires that gains from adjacent sky sections have similar values \citep{Weeren2016}.
\end{enumerate}

Direction-dependent calibration can be a good calibration approach when antenna responses are difficult to model. The direction-dependent gains can fit response shapes that are not well-modeled and vary antenna-to-antenna. Additionally, direction-dependent calibration can be important for modeling ionospheric effects. This is particularly critical for large arrays where antennas have different lines-of-sight through the ionosphere. Refraction through the ionosphere imposes direction-dependent image distortions that vary in time as ionospheric conditions change \citep{Jordan2017, Albert2020}.

Direction-dependent, per-frequency calibration increases the calibration degrees-of-freedom in a way that can suppress diffuse foreground emission and the cosmological signal itself, leading to signal loss. One can protect against this signal loss by reducing calibration degrees-of-freedom across frequency \citep{Patil2016}. In \S\ref{s:freq_cal} we explore alternatives to per-frequency calibration. With any calibration approach, one must quantify and mitigate potential signal loss stemming from coupling of the tunable calibration parameters and the cosmological signal.

\section{Simulation Results}
\label{s:sim_results}

In this section we present results from a simplified simulation exploring the novel calibration approaches described in \S\ref{s:general_framework}. The simulation results illustrate that a unified calibration approach that combines redundant calibration with a sky model prior can improve calibration performance.

\subsection{Simulation Setup}

\begin{figure}
\centering
\includegraphics[width=\columnwidth]{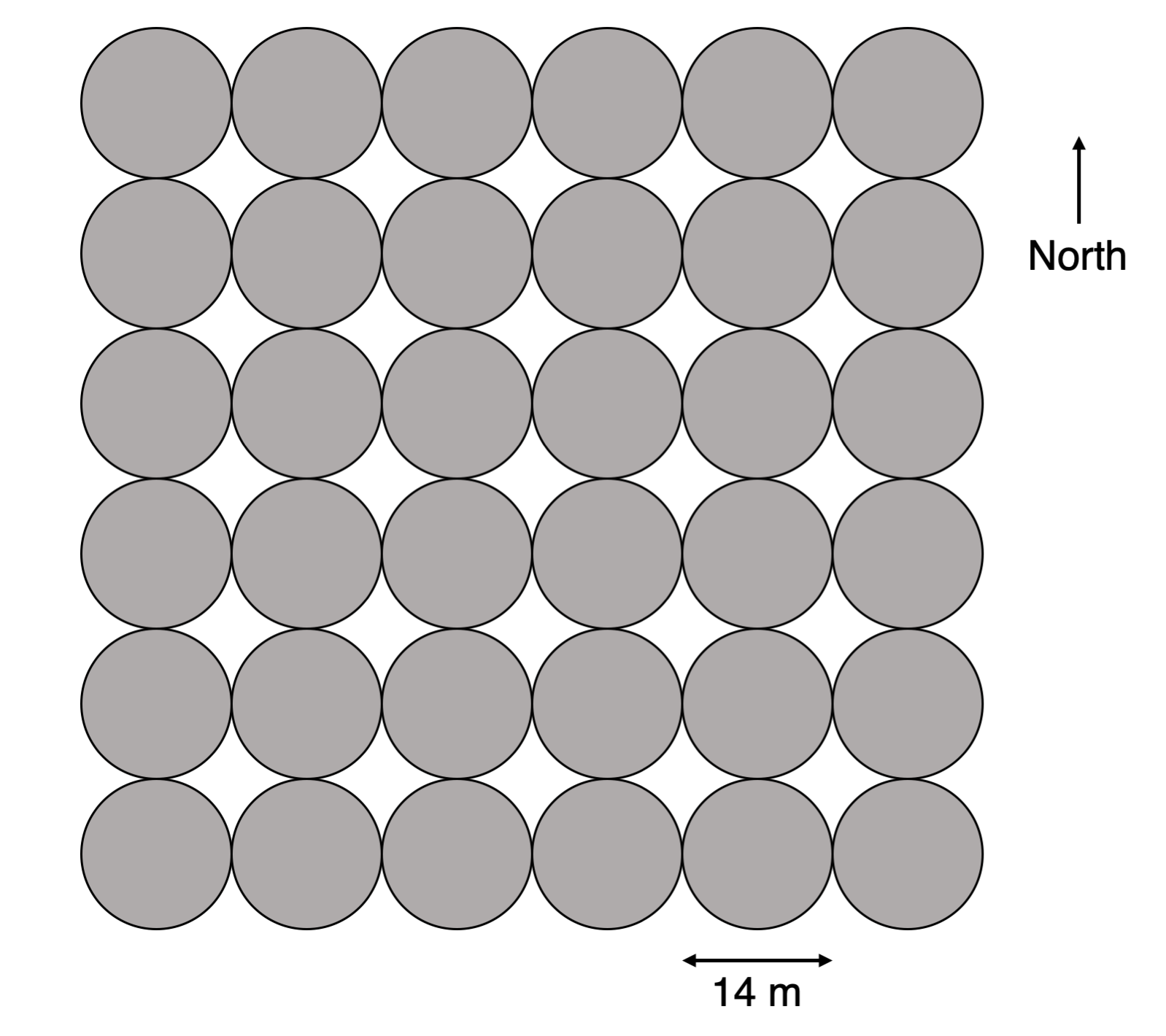}
\caption{Schematic of the array layout used in simulation. The antennas are represented as uniform circular apertures with diameters of 14 m. The array comprises 36 antennas in a close-packed regular $6 \times 6$ grid.}
\label{fig:array}
\end{figure}

The simulation is based on an idealized redundant array of 36 antennas configured as a $6\times 6$ square grid. The antennas are simulated as uniform circular apertures with diameters of 14 m and the array is close-packed: antenna centers are separated by 14 m. Figure \ref{fig:array} provides a schematic of the array configuration. We simulate visibilities from the GLEAM catalog \citep{hurley-walker2017} with the pyuvsim simulation package \citep{Lanman2019}. The simulation corresponds to a zenith observation of the MWA's `EoR-0' field, a field studied extensively by the MWA's EoR team centered at Right Ascension 0 hours and Declination $-27^\circ$.

We calibrate according to the framework presented in Equation 
\ref{eq:red_cal_most_general}. We assume perfect array redundancy and take $\fitparamsu$ to correspond to visibilities from redundant baseline sets. $\matrixa$ therefore takes the form of Equation \ref{eq:A_trad_red_cal}.

In contrast to traditional redundant calibration, this calibration implementation accounts for non-zero correlations between baselines from different redundant sets. As we model antennas as circular apertures, each antenna response is given by
\begin{equation}
    \antres(\groundcoord) \propto \begin{cases}
    1, &\text{for $|\groundcoord| < a/2$} \\
    0, &\text{otherwise}
    \end{cases},
\end{equation}
where $\groundcoord$ is the vector position on the ground from the antenna center and $a$ is the antenna diameter (14 m in our simulation).
The baseline response is evaluated by convolving the antenna responses:
\begin{equation}
    \beamvec(\uvcoord) \propto \int_{-\infty}^{\infty} \antres(\groundcoord) \antres^*(\uvcoord - \groundcoord) d^2\groundcoord,
\end{equation}
where $\uvcoord$ is the vector distance from the baseline center in the \textit{uv} plane.
Evaluating this integral and normalizing such that the integrated baseline response is equal to 1, we get that
\begin{equation}
    \beamvec(\uvcoord) =
    \frac{8}{\pi^2 a^2} \left[ \cos^{-1} \left(\frac{|\uvcoord|}{a}\right) - \frac{|\uvcoord|}{a} \sqrt{1-\left(\frac{|\uvcoord|}{a}\right)^2} \right]
\end{equation}
for $|\uvcoord| \le a$.

Following Equation \ref{eq:cov_calc_integral}, we define the baseline correlation matrix $\modelcorr$ as
\begin{equation}
    \modelcorr{}_{,jk} = \operatorname{corr}[\fitparamsu_j, \fitparamsu_k^*] = \frac{\int_{-\infty}^\infty \beamvec(\uvcoord) \beamvec(\uvcoord-\Delta \uvcoord_{jk}) d^2\uvcoord}{\int_{-\infty}^\infty \beamvec^2(\uvcoord) d^2\uvcoord}
\label{eq:baseline_corr}
\end{equation}
where $\Delta \uvcoord_{jk}$ is the \textit{uv} separation between the centers of baselines $j$ and $k$. From the definition of the correlation, the matrix is normalized such that the diagonal elements $\modelcorr{}_{,jj} = 1$. For our array, the closest non-redundant baselines are separated by 14 m. Numerically evaluating the integrals in Equation \ref{eq:baseline_corr}, we find that those baselines have correlations of 0.1617. The next closest baseline separations are 19.80 m; those baselines have correlations of 0.0176. Subsequently more distant baselines have no \textit{uv} overlap and therefore no correlation. We define the model covariance matrix from these correlation values: $\modelcov = \sigma_\text{M}^2 \modelcorr$, where $\sigma_\text{M}^2$ quantifies the variance on the sky model errors. The inverse quantity $\modelcov^{-1} = \frac{1}{\sigma_\text{M}^2} \modelcorr^{-1}$. $\modelcorr$ is full-rank; we invert it with the numpy linear algebra package.

Combining these elements, we produce $\negloglikelihood(\gains,\fitparamsu)$ for calibration of the form 
\begin{equation}
\begin{split}
    \negloglikelihood(\gains,\fitparamsu) = & \frac{1}{\sigma_\text{T}^2} \sum_{ab} \sum_j \left|\data_{ab} - \gains_a \gains_b^* \matrixa_{ab,j} \fitparamsu_j \right|^2 \\
    &+ \frac{1}{\sigma_\text{M}^2} (\fitparamsu - \modelvals)^\dag \modelcorr^{-1} (\fitparamsu - \modelvals).
\end{split}
\label{eq:cal_sim}
\end{equation}
As explained in \S\ref{s:general_framework}, in the limit that $\sigma_\text{M}^2 \rightarrow \infty$ the second term of $\negloglikelihood(\gains,\fitparamsu)$ disappears and we recover traditional redundant calibration. In the opposite limit that $\sigma_\text{M}^2 \rightarrow 0$, $\fitparamsu \rightarrow \modelvals$ and we recover traditional sky-based calibration. We can also operate between these limits by setting $\sigma_\text{M}^2$ to a finite value that represents our confidence in the model visibilities $\modelvals$.

Unified calibration has one degenerate parameter, corresponding to the overall phase of the gains. We constrain this term by requiring that the average complex phase of the gains is zero.

To simulate thermal noise, we inject an independent random signal into the each of the 630 simulated visibilities $\data$. The injected noise is drawn from a complex Gaussian distribution with a variance of 0.04 Jy$^2$; reflecting that, we set $\sigma_\text{T}^2 = 0.04 \text{ Jy}^2$. We explore two distinct classes of sky model errors. In \S\ref{s:random_error_sim} we introduce random error into the model visibilities $\modelvals$, whereas in \S\ref{s:incomplete_model_sim} we introduce errors in $\modelvals$ by omitting faint sources from the sky model. In both cases the variance on the model visibility errors is 0.16 Jy$^2$, so we set $\sigma_\text{M}^2 = 0.16 \text{ Jy}^2$.

\subsection{Simulating Random Model Error}
\label{s:random_error_sim}

In this simulation, we represent error on the model visibilities $\modelvals$ by injecting an independent random signal into each of the 60 unique baseline measurements. The error is drawn from a complex Gaussian distribution with a variance of 0.16 Jy$^2$. This technique allows us to explore the implications of sky model error across a statistical ensemble of error realizations. In \S\ref{s:incomplete_model_sim} we examine a more physical case in which error stems from faint missing sources in the sky model.

\begin{figure*}
\centering
\includegraphics[width=7in]{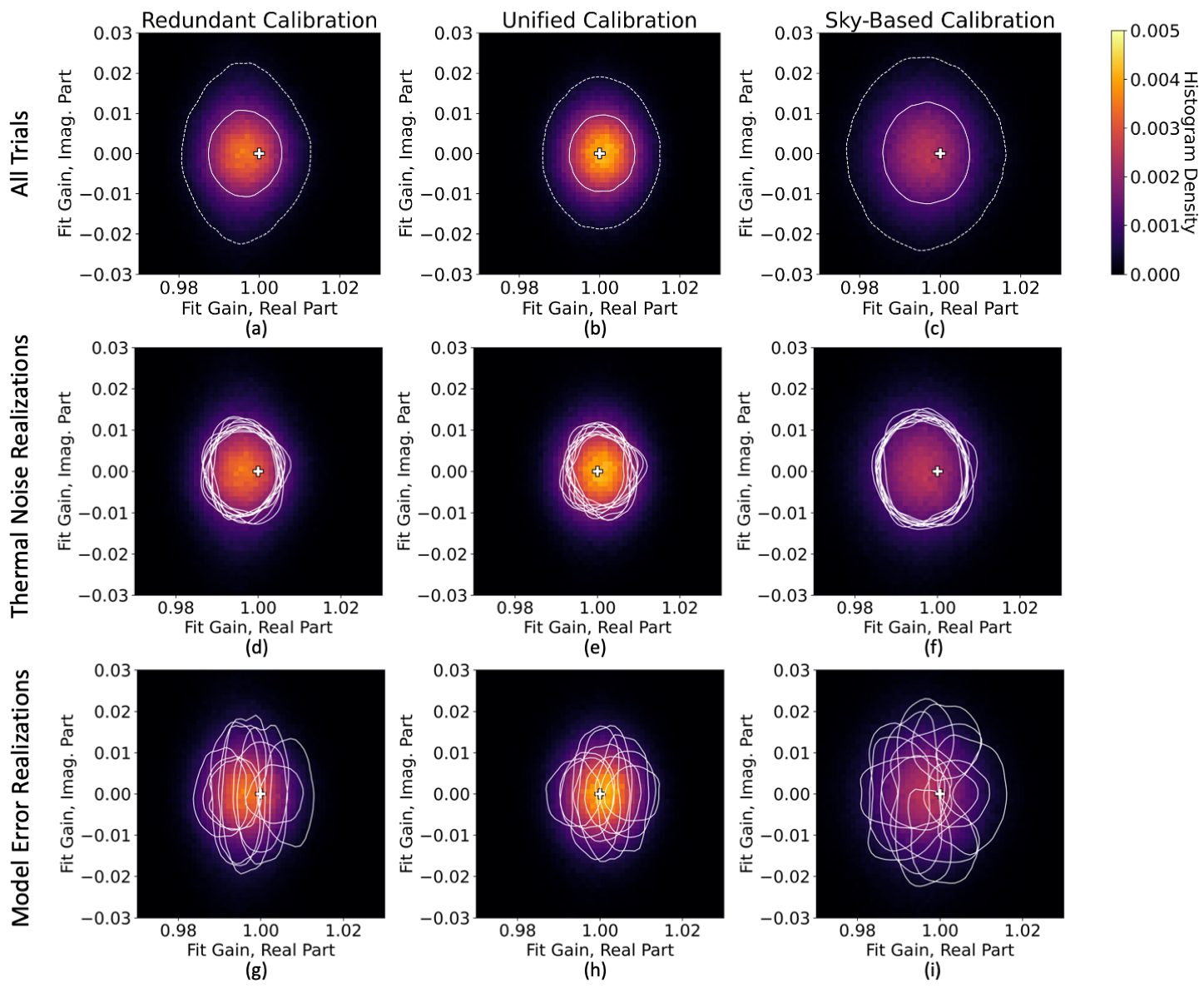}
\caption{2-D histograms of the fit gains $\hat{\gains}$ from a simulation of the array depicted in Figure \ref{fig:array} in the presence of random thermal noise and sky model error. The true gains are 1 (marked by a white plus sign), and deviations from 1 correspond to errors in the fit values. The three columns show the results of three different classes of calibration: traditional redundant calibration on the left, unified calibration implemented by minimizing Equation \ref{eq:cal_sim} in the middle, and traditional sky-based calibration on the right. The histograms depict 10,000 calibration trials corresponding to 100 separate realizations of injected thermal noise and 100 realizations of sky model error. Each row depicts identical histograms with different contours overlaid. For the top row, the white solid and dashed contours enclose an estimated 50\% and 90\% of the full distribution, respectively. We see that unified calibration performs better than either traditional redundant or sky-based calibration as the histogram distribution is better localized to the correct value of $\hat{\gains}=1$. In the middle row, each of the 10 white contours corresponds to a single realization of thermal noise. Each contour encloses 50\% of the distribution of fit gains from that noise realization. In the bottom row, each of the 10 contours corresponds to a single realization of model error, again enclosing 50\% of the distribution from that realization. This demonstrates that the distribution of fit gains is highly sensitive to the sky model error.}
\label{fig:sim_results_random_model}
\end{figure*}

We implement calibration by minimizing Equation \ref{eq:cal_sim} with a scipy optimization routine. We perform 10,000 trials, corresponding to 100 independent realizations of each thermal noise and sky model error. Histograms of the calibrated gains $\hat{\gains}$ over all 10,000 trials are plotted in the center column of Figure \ref{fig:sim_results_random_model}. The true gains $\hat{\gains}=1$, so deviations from 1 represent errors in the fit gains.

In addition, we calibrate the same noisy data and sky models using traditional redundant and sky-based calibration. The results from redundant calibration are plotted in the left column of Figure \ref{fig:sim_results_random_model} while the results from sky-based calibration are plotted in the right column. 

Of the three classes of calibration, we find that unified calibration performs best, producing the least gain error. The distribution of fit gains $\hat{\gains}$ is more compact for unified calibration than for either redundant or sky-based calibration. Furthermore, its peak is better localized to $\hat{\gains}=1$. In redundant and sky-based calibration, the gain amplitudes are systematically biased low. In redundant calibration this feature results from the absolute calibration step, which constrains the overall amplitude of the gains. The bias is consistent with previous results in the field \citep{Byrne2019} and stems from decoherence of the data and model visibilities in the presence of model errors. We explore this effect in more depth in Appendix \ref{app:gain_amps}.

The three rows in Figure \ref{fig:sim_results_random_model} depict the same underlying histograms with different overlaid contours. The contours in the top row highlight the distribution of the full set of 10,000 calibration trials. The solid white contour and the dashed white contour enclose an estimated 50\% and 90\% of the distribution, respectively, as calculated from a Gaussian Kernel Density Estimator (KDE). 

In the second row of Figure \ref{fig:sim_results_random_model} each contour describes the distribution of fit gains derived from a single realization of thermal noise. The contours enclose an estimated 50\% of the distribution calculated across all 100 realizations of model error. We plot 10 such contours, randomly selected from the 100 total realizations of thermal noise. While there is some variation between the contours, the variation is small. Each realization of thermal noise produces a distribution of fit gain errors representative of the overall distribution.

In contrast, in the bottom row of Figure \ref{fig:sim_results_random_model} each contour corresponds a single realization of sky model error. These contours are calculated to enclose 50\% of the distribution of fit gains across all 100 realizations of thermal noise. We see that independent draws of sky model error produce very different gain distributions. This is because model error is coherent across measurements within a redundant baseline set. As regular arrays have few independent measurements of the sky signal, sky model error affecting just one baseline mode can impact calibration for a large fraction of the measured visibilities. Calibration is typically implemented for observations of a single field on the sky, so in practice we work with a single realization of sky model error. Unlike with thermal noise, the impact of sky model error on calibration solutions will not average down across long data integrations.

\begin{figure*}
\centering
\includegraphics[width=4.84in]{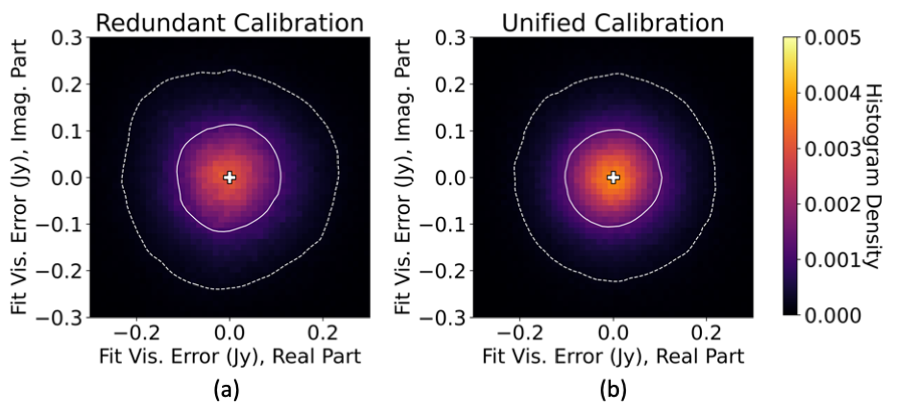}
\caption{Complex error on fit visibility parameters $\hat{\fitparamsu}$ from the simulation presented in Figure \ref{fig:sim_results_random_model}. (a) presents the results from traditional redundant calibration and (b) presents the results from unified calibration. Traditional sky-based calibration does not fit visibility parameters. The white plus signs mark the position of zero error, and the white solid and dashed contours enclose an estimated 50\% and 90\% of the data, respectively. Unified calibration produces slightly more precise values of $\hat{\fitparamsu}$ than traditional redundant calibration.}
\label{fig:sim_vis_random_model}
\end{figure*}

We can likewise compare the fit visibility values $\hat{\fitparamsu}$ between traditional redundant calibration and unified calibration. Figure \ref{fig:sim_vis_random_model} presents histograms of the error in $\hat{\fitparamsu}$. Figure \ref{fig:sim_vis_random_model}a corresponds to traditional redundant calibration and Figure \ref{fig:sim_vis_random_model}b corresponds to unified calibration. Traditional sky-based calibration does not fit visibility values. Figure \ref{fig:sim_vis_random_model} demonstrates that unified calibration produces slightly less fit visibility error than redundant calibration, as the distribution is marginally more compact. That said, the errors in the fit gains $\hat{\gains}$ are more relevant for evaluating calibration performance because the gains are applied to the data in calibration.

These simulations illustrate that, in a simplified case, the calibration framework presented in \S\ref{s:general_framework} improves upon traditional calibration approaches. Traditional sky-based calibration does not fit visibility values $\fitparamsu$ and instead only uses the model visibilities $\modelvals$, which are susceptible to errors from the sky model. Unified and redundant calibration approaches introduce additional calibration degrees-of-freedom through $\fitparamsu$ and thereby allow calibration to fit out sky model errors. 

Redundant calibration uses the model visibilities $\modelvals$ for absolute calibration only. It fits the relative calibration parameters entirely from the data and imposing agreement between redundant measurements. However, thermal noise introduces error into the relative calibration parameters. Unified calibration uses the model visibilities as a prior on calibration solutions. The prior is weighted by the measured uncertainty on the sky model. This further constrains redundant calibration's relative calibration parameters and mitigates error from thermal noise.

The calibration simulations presented in this section illustrate one implementation style of the calibration framework described in \S\ref{s:general_framework}. The simulated instrument is perfectly redundant and we assume no errors in the instrument model. Instead, calibration errors result only from error in the model visibilities $\modelvals$ and from thermal noise. The results in Figure \ref{fig:sim_results_random_model} demonstrate that, for our simplified simulation, using a unified calibration framework based on a physically-motivated instrument model produces better calibration results than traditional methods. In the next section, we demonstrate that unified calibration similarly improves calibration performance in the case that sky model errors stem from faint missing sources.

\subsection{Simulating Model Incompleteness}
\label{s:incomplete_model_sim}

In \S\ref{s:random_error_sim} we explored calibration across an ensemble of random sky model errors. As a more physically-motivated simulation, we explore the case that sky model errors stem from missing faint missing sources. This technique follows approaches used in \citealt{Barry2016} and \citealt{Byrne2019} and represents the fact that real source catalogs always have a completeness limit dictated by the measurement sensitivity.

\begin{figure}
\centering
\includegraphics[width=\columnwidth]{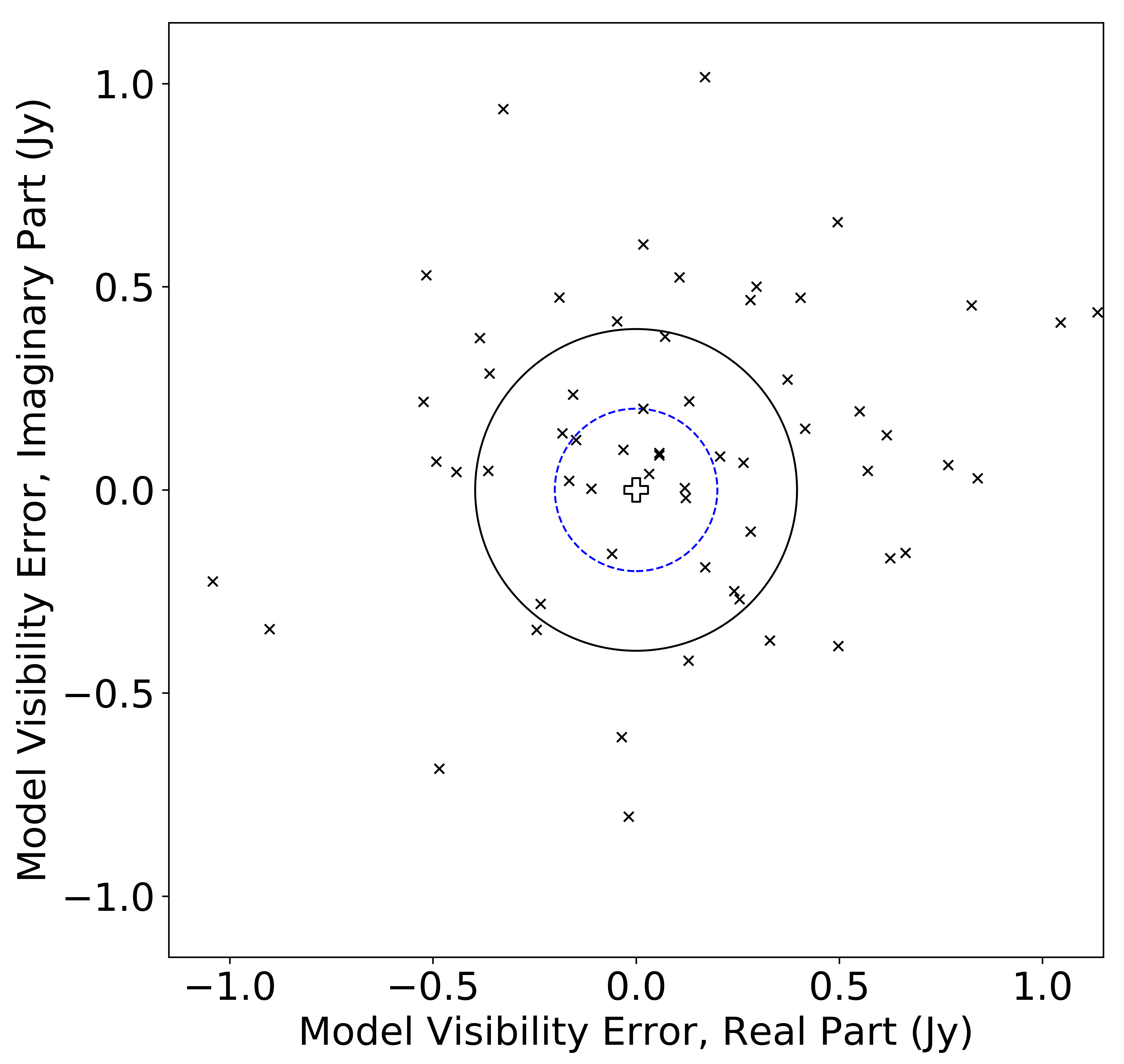}
\caption{Scatter plot of the errors in the simulated model visiblities $\modelvals$ (denoted with black `X' symbols). Data visibilities are simulated with the full GLEAM catalog and have an average amplitude of 5.58 Jy; model visibilities are simulated from only the GLEAM sources brighter than 100 mJy. The errors in the model visibilities stem from these missing faint sources. The white plus sign marks the position of zero error. The black solid contour has a radius equal to the standard deviation of the model visibility errors. The blue dashed contour has a radius equal to the standard deviation of the complex Gaussian distribution from which we simulate thermal noise.}
\label{fig:model_vis}
\end{figure}

To represent this source of model error we once again simulate visibilities from the GLEAM catalog, this time omitting sources with flux densities less than 100 mJy. We take these simulated visibilities to be the model visibilities $\modelvals$. The missing faint sources introduce disparities between $\modelvals$ and the visibilities simulated with the full catalog (plotted as black `X' symbols in Figure \ref{fig:model_vis}). The variance of these disparities is 0.16 Jy$^2$, compared to the average visibility amplitude of 5.58 Jy. The black circular contour in Figure \ref{fig:model_vis} has a radius equal to the standard deviation of the errors.

\begin{figure*}
\centering
\includegraphics[width=7in]{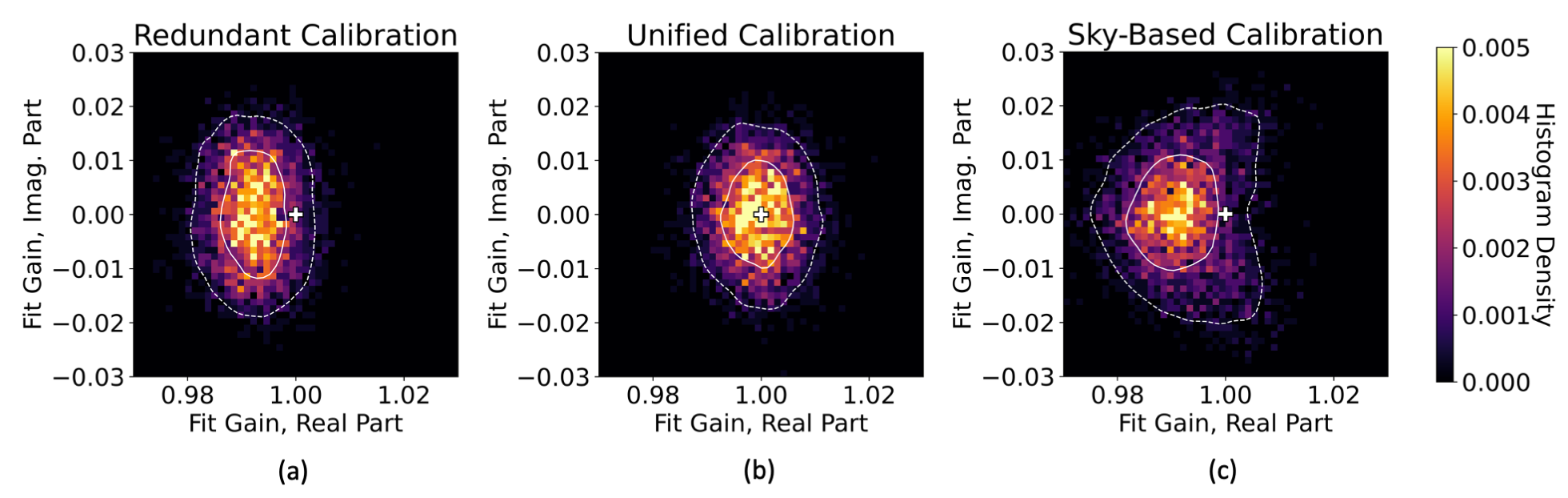}
\caption{2-D histograms of the fit gains $\hat{\gains}$ from a simulation of the array depicted in Figure \ref{fig:array}. Sky model error is represented by excluding faint sources in the simulated catalog. The true gains are 1 (marked by a white plus sign), and deviations from 1 correspond to errors in the fit values. The three plots show the results of three different classes of calibration: (a) redundant calibration, (b) unified calibration, and (c) sky-based calibration. The white solid and dashed contours enclose an estimated 50\% and 90\% of the data, respectively. The histogram distribution in (b) is better localized to the zero error point than either (a) or (c), indicating that unified calibration delivers better calibration solutions than either traditional redundant or sky-based calibration.}
\label{fig:sim_results}
\end{figure*}

As in \S\ref{s:incomplete_model_sim}, we calibrate across 100 independent realizations of injected thermal noise drawn from a complex Gaussian distribution. The resulting calibrated gains $\hat{\gains}$ are presented as histograms in Figure \ref{fig:sim_results}. Figure \ref{fig:sim_results}a corresponds to traditional redundant calibration, Figure \ref{fig:sim_results}b corresponds to unified calibration, and Figure \ref{fig:sim_results}c corresponds to sky-based calibration. As in Figure \ref{fig:sim_results_random_model}, we find that unified calibration performs better than each redundant or sky-based calibration. The fit gain distribution is most compact for unified calibration, and redundant and sky-based calibration produce gains with amplitudes that are biased low.

\begin{figure*}
\centering
\includegraphics[width=4.84in]{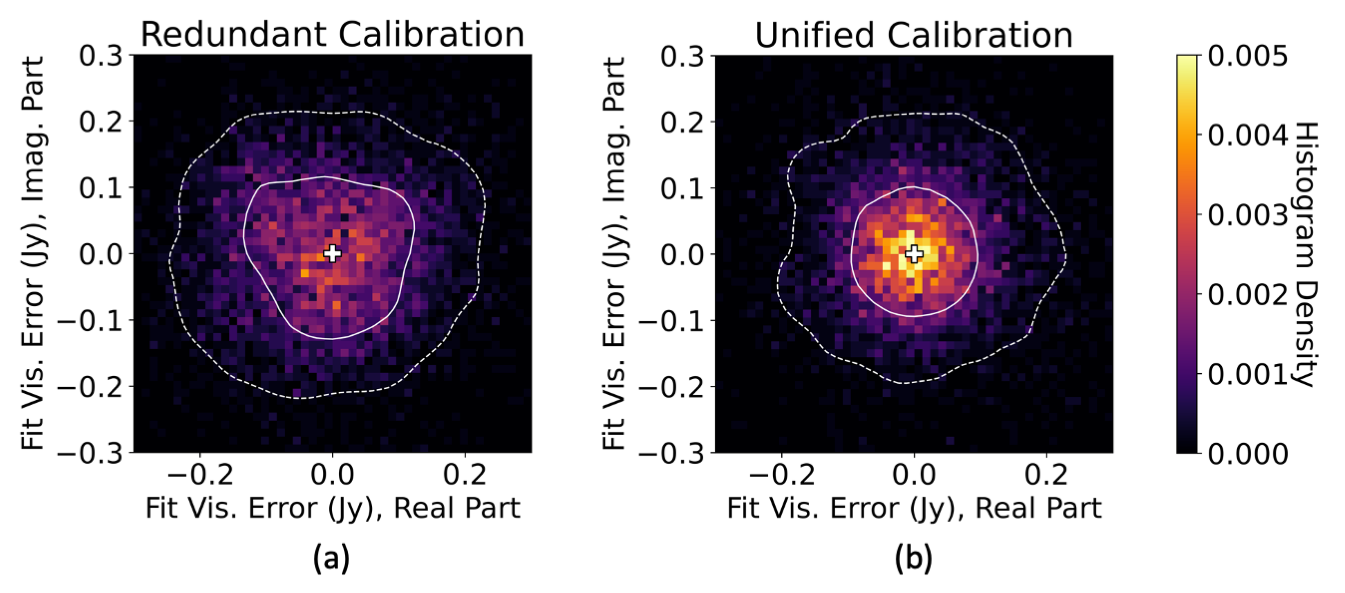}
\caption{Complex error on the fit visibility parameters $\hat{\fitparamsu}$ from the simulation presented in Figure \ref{fig:sim_results} for redundant calibration (a) and unified calibration (b). The white solid and dashed contours enclose an estimated 50\% and 90\% of the data, respectively.}
\label{fig:sim_results_vis}
\end{figure*}

In Figure \ref{fig:sim_results_vis} we plot histograms of the error on the fit visibility values $\hat{\fitparamsu}$ for redundant and unified calibration. As in Figure \ref{fig:sim_vis_random_model}, we find that unified calibration produces less fit visibility error than redundant calibration.

The results of this simulation indicate that, just as unified calibration improved calibration in the presence of the random model error simulated in \S\ref{s:random_error_sim}, it can reduce calibration error in the more physical scenario of missing faint sources in the sky model. The unified calibration framework described by Equation \ref{eq:cal_sim} assumes that the error on the model visibilities $\modelvals$ is Gaussian. This could be a poor approximation for some classes of sky model errors, for example, if the sky model error is dominated by mismodeling of a single bright source. Further work could develop more accurate models of sky model error based on specific knowledge of the sky. However, we expect that this assumption is appropriate for the case of sky model error from many faint missing sources distributed across the sky (and Figure \ref{fig:model_vis} gives no indication that the errors on $\modelvals$ are dramatically non-Gaussian). The results in this section show that the calibration framework represented by Equation \ref{eq:cal_sim} can improve upon traditional calibration methods in the presence of missing sky model sources.


\section{Frequency Calibration}
\label{s:freq_cal}

In the previous sections we implicitly assume a calibration likelihood function that is separable in frequency:
\begin{equation}
    P(\fitparams|\data) = \prod_f P_f[\fitparams(f)|\data(f)].
\end{equation}
A likelihood function of this form allows for each frequency to be calibrated separately but precludes the calibration model from incorporating cross-frequency covariance information. In this section we describe calibration extensions that break the assumption of separability of the likelihood function in frequency.

Frequency-dependent calibration errors pose a major problem for 21 cm cosmology. In particular, fast frequency structure in the calibrated gains can contaminate the cosmological signal by limiting the separability of the cosmological signal and the astronomical foregrounds.

Direction-independent gain-based calibration assumes that the instrumental response is well-modeled as a multiplicative factor on the true sky signal. In other words, a visibility is given by
\begin{equation}
    \data_{ab}(f) \approx \boldsymbol{G}_{ab}(f) \fitparamsu_{ab}(f)
\label{eq:meas_eq}
\end{equation}
where $\fitparamsu_{ab}(f)$ is the true sky signal at frequency $f$ and $\boldsymbol{G}_{ab}(f)$ is the instrumental gain. If calibration is error-free, one can divide the data by the gains to recover the true sky signal $\fitparamsu_{ab}(f)$, which contains both the cosmological signal and contamination from foreground emission. 

The basis of 21 cm cosmology is the principle that these signals are spectrally distinguishable. Foreground contamination is assumed to be spectrally smooth and can therefore be separated from the specrally variant cosmological signal. If we use $\fitparamsu_{ab}(\eta)$ to represent the Fourier transform of the signal --- where $\eta$ is the Fourier dual of frequency, or delay, with units time --- we expect $\fitparamsu_{ab}(\eta)$ to be free from foreground contamination at large $\eta$ . The precise contaminated modes depend on the baseline length, and this contaminated region is often called the `wedge' \citep{Furlanetto2006,Datta2010,Morales2010,Vedantham2012,Morales2012,Parsons2012,Trott2012,Dillon2013,Hazelton2013,Thyagarajan2013,Thyagarajan2015}.

From the convolution theorem we can rewrite Equation \ref{eq:meas_eq} as
\begin{equation}
    \data_{ab}(\eta) \approx \boldsymbol{G}_{ab}(f) * \fitparamsu_{ab}(\eta)
\end{equation}
where $*$ denotes the convolution. Now consider the case that the gains calculated in calibration have an error that contributes to a particular delay mode $\eta_0$. For that mode, $\boldsymbol{G}_{ab}(\eta_0) \rightarrow \boldsymbol{G}_{ab}(\eta_0) + \epsilon_{ab}(\eta_0)$, where $\epsilon_{ab}(\eta_0)$ is the error on the gains. Now calibration cannot recover the true sky signal; instead, it recovers the signal $\epsilon_{ab}(\eta) * \fitparamsu_{ab}(\eta)$. The error on the gains therefore convolves all the true sky power, including the foreground power. If $\eta_0$ is large, this means that foreground power is coupled into the modes we expect to be free from contamination, otherwise known as the `EoR window.' This contaminates the cosmological signal and can preclude a detection.

For this reason, it is critical that calibration is essentially free from errors with fast frequency structure. These errors can take two forms. First of all, we require that calibration does not introduce new fast frequency structure by overfitting noise. This noise could correspond to thermal noise, which would be mitigated by extending the integration time. More importantly, sidelobe noise from errors in the calibration sky model introduces fast frequency structure that does not average down in time. Secondly, we require that calibration accurately captures any true frequency structure in the instrumental response.

\subsection{Parameterizing Gains Across Frequency}
\label{s:cross_freq_gain_params}

One approach to eliminating frequency-dependent calibration errors is to fit calibration solutions across frequency after-the-fact \citep{Barry2019, Li2019}. With this approach, calibration consists of two steps. First, one calculates per-frequency calibration solutions. Next, the gains are adjusted to remove false frequency structure and fit the allowable degrees of freedom determined by a model of the instrument bandpass response.

A more optimal way of calibrating with a modeled bandpass is to fit the parameters of the bandpass directly \citep{Yatawatta2015, Mertens2020}. We can model the gains as functions of tunable parameters $\freqparams_a$ for each antenna $a$ such that
\begin{equation}
    \gains_a(f) = \gains_a(f,\freqparams_a).
\end{equation}
Here $\freqparams_a$ could include the amplitudes and positions of known features in the bandpass or any other parameters related to fitting the instrument's bandpass response. If the bandpass is parameterized as a low-order polynomial, $\freqparams_a$ could be the polynomial coefficients:
\begin{equation}
    \gains_a(f,\freqparams_a) = \sum_{n=0}^{n_\text{max}} \freqparams_{a,n} f^n,
\end{equation}
where $n_\text{max}$ is the maximum mode in the low-order polynomial fit. 

Because the parameters $\freqparams_a$ are not per-frequency, calibration can no longer be parallelized across frequency channels. Instead, the algorithm must access data from across a large frequency range at once. This increases the calibration memory requirements by a factor equal to the number of frequency channels, which can be computationally infeasible. Novel calibration algorithms allow for distributed optimization without sacrificing cross-frequency performance \citep{Yatawatta2015}.

An extension to traditional sky-based calibration (Equation \ref{eq:sky_cal}) that parameterizes the gains across frequencies has the following features:
\begin{enumerate}
    \item We define tunable calibration parameters corresponding to parameterizations of the antenna gains across frequency. $\fitparams = \{ \freqparams_1, \freqparams_2, \dots \}$ where for antenna $a$ the gain $\gains_a(f) = \gains_a(f,\freqparams_a)$.
    \item We model the data as 
    \begin{equation}
        \visfunc_{ab}(f)=\gains_a(f,\freqparams_a) \gains_b^*(f,\freqparams_b) \modelvals_{ab}(f)
    \end{equation}
    where indices $a$ and $b$ index antennas.
    \item As in previous sections, we describe $P(\data|\fitparams)$ as an independent Gaussian probability distribution.
    \item We use a flat prior such that $P(\fitparams)=P(\freqparams_1, \freqparams_2, \dots)$ is a constant.
\end{enumerate}

This version of sky-based calibration has a negative log-likelihood of the form
\begin{equation}
\begin{split}
    & \negloglikelihood(\freqparams_1, \freqparams_2, \dots) = \sum_f \sum_{ab} \frac{1}{\thermalvar{}_{,ab}(f)} \\
    & \times \left|\data_{ab}(f) - \gains_a(f,\freqparams_a) \gains_b^*(f,\freqparams_b) m_{ab}(f)\right|^2.
\end{split}
\end{equation}
Calibration consists of minimizing this quantity by varying the parameters $\freqparams_a$ for each antenna $a$.

This calibration approach allows for variable weighting across frequency. For example, frequency channels with greater noise contamination could be downweighted by increasing $\thermalvar(f)$ with respect to the uncontaminated channels. Frequency channels can be removed from calibration altogether by taking the limit $\thermalvar(f) \rightarrow \infty$. If the calibration parameters are chosen to span frequencies then they will interpolate the gains across the masked frequency channels.

This approach to fitting the gains is not only applicable to sky-based calibration. One can choose to parameterize the gains across frequency in any calibration framework.

\subsection{Capturing Redundancy Across Frequency}

Calibrating across frequencies may allow redundant calibration to capture cross-frequency visibility covariances (Dillon \& Parsons, private communication). It is possible that baselines of vastly different physical lengths are nonetheless highly covariant at different frequencies. A long baseline could measure the same sky modes at a low frequency that a short baseline measures at a higher frequency. This has interesting implications for redundant calibration. As in \S\ref{s:cross_freq_gain_params}, this calibration approach would pose computational challenges as it precludes parallelization across frequency channels.

To illustrate cross-frequency redundant calibration we consider an extension to redundant calibration as described in \S\ref{s:red_cal}. Under this extension, calibration includes the following assumptions:
\begin{enumerate}
    \item As in \S\ref{s:trad_cal} and \S\ref{s:general_framework}, we parameterize the gains as per-antenna, per-frequency, and direction-independent. (Alternatively, one could combine this calibration approach with the cross-frequency gain parameterizations described in \S\ref{s:cross_freq_gain_params}. However, for simplicity we will describe a per-frequency gain parameterization.) We parameterize the sky model with per-frequency visibilities $\fitparamsu(f)$. The tunable calibration parameters are $\fitparams=\{ \gains(f), \fitparamsu(f)\}$.
    \item As in \S\ref{s:general_framework}, we model the data as
    \begin{equation}
        \visfunc_{ab}(f)= \gains_a(f) \gains_b^*(f) \sum_j \matrixa_{ab, j} \fitparamsu_j(f)
    \end{equation}
    for some matrix $\matrixa$ that maps $\fitparamsu$ to visibilities.
    \item As before, $P(\data|\fitparams)$ is an independent Gaussian probability distribution (Equation \ref{eq:independent_gaussian}).
    \item We use a Gaussian prior on $\fitparamsu$ and a flat prior on $\gains$ such that
    \begin{equation}
    \begin{split}
    &P(\fitparams) \propto P(\fitparamsu) \propto \\ 
    & \prod_{f_1 f_2} e^{-\frac{1}{2} [\fitparamsu(f_1) - \modelvals(f_1)]^\dag \mathbf{C}_{M, f_1 f_2}^{-1} [\fitparamsu(f_2) - \modelvals(f_2)]}.
    \end{split}
    \end{equation}
    $\modelcov$ now encodes covariances between every visibility at every frequency, with elements given by
    \begin{equation}
        \modelcov{}_{,j k f_1 f_2} = \operatorname{cov}[\fitparamsu_j(f_1), \fitparamsu_k^*(f_2)]
    \end{equation}
    where $j$ and $k$ index baselines and $f_1$ and $f_2$ index frequency channels.
\end{enumerate}

$\negloglikelihood(\gains,\fitparamsu)$ from Equation \ref{eq:red_cal_most_general} now takes the form
\begin{equation}
\begin{split}
    & \negloglikelihood(\gains,\fitparamsu) = \sum_f \sum_{ab} \sum_j \frac{1}{\sigma_{T,ab}^2(f)}\\
    & \qquad \times \left|v_{ab}(f) - \gains_a(f) \gains_b^*(f) \matrixa_{ab,j} \fitparamsu_j(f)\right|^2 \\
    &+ \sum_{f_1 f_2} \sum_{jk} [\fitparamsu_j(f_1) - \modelvals_j(f_1)]^* \modelcov{}_{,j k f_1 f_2}^{-1} [\fitparamsu_k(f_2)-\modelvals_k(f_2)].
\end{split}
\end{equation}
This converges to traditional redundant calibration in the limit that $\modelcov{}_{,j k f_1 f_2} \rightarrow 0$ for $f_1 \neq f_2$.

Assuming a stable sky as a function of frequency, a physically-motivated construction of $\modelcov$ could follow the approach described in \S\ref{s:red_cal_known_imperfect_redundancy}. From Equation \ref{eq:cov_calc_integral} we get
\begin{equation}
\begin{split}
    \modelcov{}_{,jk f_1 f_2} =  & \int_{-\infty}^\infty \int_{-\infty}^\infty \beamvec_j(f_1, \uvcoord) \beamvec_k^*(f_2, \uvcoord') \\
    & \times \operatorname{cov}[\boldsymbol{S}(f_1, \uvcoord), \boldsymbol{S}^*(f_2, \uvcoord')] d^2\uvcoord d^2\uvcoord',
\end{split}
\end{equation}
where $\beamvec_j(f, \uvcoord)$ is the \textit{uv} response of baseline $j$ and $\boldsymbol{S}(f, \uvcoord)$ is the \textit{uv} plane at frequency $f$. If, as in \S\ref{s:red_cal_known_imperfect_redundancy}, we assume that different points in the \textit{uv} plane are independent and that the variance across the \textit{uv} plane is constant, we get that
\begin{equation}
    \modelcov{}_{,jk f_1 f_2} = \boldsymbol{\mathsf{C}}_{\text{S},f_1 f_2} \int_{-\infty}^\infty \beamvec_j(f_1, \uvcoord) \beamvec_k^*(f_2, \uvcoord) d^2\uvcoord.
\label{eq:vis_freq_cov}
\end{equation}
Here $\boldsymbol{\mathsf{C}}_{\text{S},f f}$ is the variance of \textit{uv} pixels at frequency $f$; $\boldsymbol{\mathsf{C}}_{\text{S},f_1 f_2}$ is the covariance of \textit{uv} pixels at frequencies $f_1$ and $f_2$.

It follows from Equation \ref{eq:vis_freq_cov} that the construction of a physically-motivated cross-frequency $\modelcov$ requires accurate modeling of the beam responses at all frequencies in question. In practice, beam responses vary considerably as a function of frequency. While baselines of different physical lengths may sample the same \textit{uv} locations at different frequencies, we cannot expect their beam responses to be the same. Therefore, redundancy is degraded across frequencies. In other words, we expect the largest elements of $\modelcov{}_{,f_1 f_2}$ to occur when $f_1 = f_2$.

If we assume the sky is constant across our frequency range then $\boldsymbol{\mathsf{C}}_{\text{S},f_1 f_2} = \boldsymbol{\mathsf{C}}_{\text{S},f_1 f_1}$ at all frequencies. However, in practice the sky has some frequency-dependence. We therefore expect that a more accurate model sets $\boldsymbol{\mathsf{C}}_{\text{S},f_1 f_2} < \boldsymbol{\mathsf{C}}_{\text{S},f_1 f_1}$ when $f_1 \neq f_2$. This further suppressed the elements of $\modelcov$ that correspond to covariances of visibilities at different frequencies.

Incorporating nonzero cross-frequency visibility covariances in calibration could be a powerful tool for constraining the frequency structure of the instrument response. However, beam shape variation across antennas and frequencies, along with frequency-dependence in the sky signal, suppress the covariance of visibilities at different frequencies. A realistic calibration model must account for these effects.

\section{Polarized Calibration}
\label{s:pol_cal}

Polarized interferometry can be accomplished when each antenna has two polarization modes. This translates to four measured polarization modes from each baseline. The polarized measurement equation is given by
\begin{equation}
    \vismat_{ab} = \sum_k \jonesmat_{ak} \coherency_k \jonesmat_{bk}^\dag + \boldsymbol{\mathsf{N}}_{ab},
\label{eq:pol_measurement}
\end{equation}
where $k$ indexes positions on the sky. Here $\vismat_{ab}$ is a $2\times2$ matrix with elements corresponding to the four polarization modes measured by baseline $\{a, b\}$. The off-diagonal elements of $\vismat_{ab}$ correspond to the `cross-visibilities' formed by correlating measurements from different polarizations of the two antennas. $\jonesmat_{ak}$ and $\jonesmat_{bk}$ are Jones matrices associated with antennas $a$ and $b$, respectively, and $\boldsymbol{\mathsf{N}}_{ab}$ gives the noise on the measurements. $\coherency_k$ is the coherency matrix, equal to
\begin{equation}
    \coherency_k = \langle \electricfield_k \electricfield_k^\dag \rangle
\end{equation}
where $\electricfield_k$ is the electric field vector on the sky at position $k$ \citep{Hamaker1996a, Hamaker2000}.

$\coherency_k$ is diagonal for an unpolarized sky, but it does not follow that $\boldsymbol{\mathsf{V}}_{ab}$ will be diagonal, even for an instrument with orthogonal antenna polarizations. Although an antenna's two polarizations may measure orthogonal polarization modes of incident radiation from zenith (or, for a mechanically steerable antenna, from the pointing center), they will measure non-orthogonal polarization modes of off-zenith radiation. This effect is particularly pronounced for widefield instruments and fields with bright off-zenith sources. For this reason, one can expect that even unpolarized incident radiation couples into the measured cross-visibilities of an interferometer.

Equation \ref{eq:pol_measurement} is often rewritten to describe the visibilities as a vector of length 4 rather than a $2\times2$ matrix. In that notation, Equation \ref{eq:pol_measurement} becomes
\begin{equation}
    \data_{ab} = \sum_k (\jonesmat_{ak} \otimes \jonesmat_{bk}^*) \boldsymbol{s}_k + \boldsymbol{n}_{ab},
\end{equation}
where $\jonesmat_{ak} \otimes \jonesmat_{bk}^*$ is the $4\times4$ Kronecker product of the Jones matrices. $\data_{ab}$, $\boldsymbol{s}_k$, and $\boldsymbol{n}_{ab}$ are the vector representations of the visibilities, the coherency matrix, and the noise, respectively.

\subsection{Per-Polarization Calibration}

Calibration frameworks presented thus far in this paper have assumed that $\negloglikelihood(\fitparams)$ is separable in polarization. Under this assumption one can write the likelihood function as
\begin{equation}
    P(\fitparams|\data) = \prod_p P_p[\fitparams_p|\data_{pp}],
\end{equation}
where $p$ indexes the instrumental polarization modes. Here $\fitparams_p$ is the set of calibration parameters associated with instrumental polarization $p$. $\data_{ab,pp}$ is the the $(p,p)$ element of the matrix $\boldsymbol{\mathsf{V}}_{ab}$; $\data_{pp}$ is the vector formed from those elements for all baselines. Note that cross-visibilties are excluded from per-polarization calibration, which amounts to leaving potentially valuable information on the table.

We can rewrite traditional sky-based calibration (Equation \ref{eq:sky_cal}) with explicit polarization dependence as
\begin{equation}
    \negloglikelihood(\gains) = \sum_{ab} \sum_{p} \frac{1}{\sigma_{T,abp}^2} \left|\data_{ab,pp} - \gains_{a,p} \gains_{b,p}^* \modelvals_{ab,pp}\right|^2.
\end{equation}
Here $\gains_{a,p}$ is the gain of the $p$-polarization of antenna $a$ and $\modelvals_{ab,pp}$ is an estimate of the visibility $\data_{ab,pp}$. The model visibilities are produced from estimations of the sky coherency matrices $\coherency_k$ and models of the Jones matrices $\jonesmat_{ak}$. 

Per-polarization calibration has a degeneracy associated with the average phase between polarizations. One can see this degeneracy by noting that the transformation $\gains_p \rightarrow \gains_p e^{i \phi}$ does not change the form of the $\negloglikelihood(\gains)$, where $\gains_p$ is the set of all $p$-polarized gains. This degeneracy means that per-polarization calibration cannot calibrate the cross-visibilities. One therefore cannot produce polarized images from per-polarization calibration. (`Pseudo-Stokes' I and Q images do not require the cross-visibilities and can approximate true Stokes I and Q near zenith.)

\subsection{Polarized Calibration with Per-Polarization Instrumental Gains}

One approach to fully polarized calibration retains the per-polarization parameterization of the gains but incorporates the cross-visibilities. A polarized extension to traditional sky-based calibration would use $\negloglikelihood(\gains)$ of the form
\begin{equation}
    \negloglikelihood(\gains) = \sum_{ab} \sum_{pq} \frac{1}{\sigma_{T,abpq}^2} \left|\data_{ab,pq} - \gains_{a,p} \gains_{b,q}^* \modelvals_{ab,pq}\right|^2
\label{eq:pol_cal}
\end{equation}
where $p$ and $q$ index instrumental polarizations.

Because this calibration framework includes the cross-visibilities it does not have the degeneracy associated with an overall phase between polarizations. Instead, it has just one degenerate parameter corresponding to the overall phase of \textit{all} gains. Although the gains are parameterized per-polarization, this calibration approach is fully polarized and can be used to produce polarized images.

Equation \ref{eq:pol_cal} represents a polarized extension to traditional sky-based calibration (Equation \ref{eq:sky_cal}), which we have implicitly defined to be per-polarization. This polarized calibration approach could be combined with other calibration frameworks to achieve, for example, polarized redundant calibration \citep{Kohn2019}.

\subsection{Calibrating Polarization Leakage}

Per-polarization gain parameterization does not allow calibration to correct for polarization leakage. Polarization leakage occurs when power modeled in one visibility mode is measured as another \citep{Sault1996}. This can occur through a number of mechanisms. An antenna can be physically mis-aligned such that, for example, the so-called East-West polarized feed is somewhat skewed North-South. Antennas can experience cross-talk --- for example, a signal from a North-South polarized feed could get coupled into the East-West feed's signal path. Additionally, the ionosphere can Faraday rotate one polarization mode into another. Virtually all real antennas exhibit some level of polarization leakage. We can expand the calibration degrees-of-freedom to capture the effects.

To correct for polarization leakage in calibration, we can parameterize each of the per-antenna gains as a $2\times2$ matrix with two polarization indices. $\gains_{a,pr}$ is then the gain of the $p$-polarization of antenna $a$ with respect to incident $r$-polarized radiation. When $p \neq r$, $\gains_{a,pr}$ encodes the polarization leakage from mode $r$ to mode $p$. These off-diagonal elements of the gains matrices are often called `D-terms.' An extension to traditional sky-based calibration that uses that calibration approach has $\negloglikelihood(\gains)$ of the form
\begin{equation}
    \negloglikelihood(\gains) = \sum_{ab} \sum_{pq} \sum_{rs} \frac{1}{\sigma_{T,abpqrs}^2} \left|\data_{ab,pq} - \gains_{a,pr} \gains_{b,qs}^* \modelvals_{ab,rs}\right|^2.
\label{eq:pol_leakage_cal}
\end{equation}
Here $p$ and $q$ index polarization modes of the measured visibilities and $r$ and $s$ index polarization modes of the modeled visibilities.

\section{Conclusion}

Precision calibration is essential to the success of 21 cm cosmology experiments. Low-level calibration errors contaminate the cosmological signal by limiting the separability of bright astrophysical foregrounds. As current experiments approach a detection of the 21 cm power spectrum from the EoR, novel calibration approaches must achieve new levels of precision.

While a diversity of interferometric calibration techniques exist, the current paradigm in the field delineates between sky-based and redundant calibration approaches. Sky-based calibration assumes very good \textit{a priori} models of the sky and instrument response. It further assumes independent visibility measurements, neglecting visibility covariance from baseline response overlap in the \textit{uv} plane. On the other hand, redundant calibration assumes that baselines within a redundant set are perfectly redundant both in antenna positions and response shapes, that baselines from different redundant baseline sets exhibit no covariance, and that a sky model can accurately constrain the absolute calibration parameters. The literature shows that violating these assumptions introduces calibration errors that contaminate the measurement. For example, \citealt{Barry2016} shows that sky model incompleteness leads to sky-based calibration errors, \citealt{Joseph2018} and \citealt{Orosz2019} explore redundant calibration errors from antenna position offsets and response irregularities, and \citealt{Byrne2019} examines the effect of sky model errors on the absolute calibration step of redundant calibration. Without mitigation, these errors can preclude a detection of the cosmological signal. 

The standard assumptions of sky-based and redundant calibration can be inflexible to demands of calibrating real data from physical interferometers. Next-generation calibration techniques, such as those presented in \citealt{Sievers2017} and this paper, combine elements of these disparate calibration approaches. New calibration frameworks must be statistically rigorous, physically-motivated, and adaptive to the complex physical systems they model. This paper presents a highly general statistical language that clarifies the assumptions implicit in a calibration formalism. It then describes new approaches to calibration that incorporate imperfect sky models, relax redundancy requirements, and apply the ideas of redundant calibration --- namely, that baselines can be covariant with one another --- to a wider class of arrays than those that are typically redundantly calibrated.

While the calibration frameworks described in this paper are highly general, further extensions to this work could expand upon them. One such extension is explicitly time-dependent calibration, where the system's evolution in time is parameterized in the calibration model. Another avenue of exploration could involve developing new approaches for constraining the overall phase of the calibration solutions. These could incorporate gated pulsar measurements, thereby augmenting sky-based or redundant calibration variants with absolute timing information. 

Additional calibration extensions could constrain the frequency structure of the calibration solutions to mitigate spectral errors in calibration. \citealt{Barry2019a} and \citealt{Li2019} fit the gains' frequency structure from the visibility autocorrelations. This protects the calibration solutions from frequency-dependent sky model errors but limits the number of visibility measurements used in calibration. As an extension to this approach, we could require that the calibration solutions fit only frequency structure that lies outside of the `foreground wedge.' This approach could retain the benefits of calibrating to the autocorrelations while improving calibration signal-to-noise. We could combine this technique with regularized optimization approaches, similar to those described in \citealt{Yatawatta2015}, to further penalize frequency structure in the calibration solutions. This could allow calibration to fit true spectral features in the instrument's response while mitigate spectral overfitting.

21 cm cosmology is pushing the limits of precision radio interferometry. Progress in the field will require excellent systematic suppression, including the mitigation of calibration errors. Novel calibration approaches can potentially improve measurement sensitivity. These approaches expand upon existing sky-based and redundant calibration techniques, relaxing the assumptions inherent in those approaches and incorporating more physical instrument models.

\section*{Acknowledgements}

We would like to thank Henry Brinkerhoff, Ronniy Joseph, Wenyang Li, and Ian Sullivan for discussions that directly contributed to this work. We thank the anonymous reviewer who provided insightful comments that were invaluable to the development of this paper. This work was directly supported by NSF grants AST-1613855, 1506024, 1643011, and 1835421.

\section*{Data Availability Statement}

The data underlying this article will be shared on reasonable request to the corresponding author.

\bibliographystyle{mnras}
\bibliography{references}

\begin{thebibliography}{}
\makeatletter
\relax
\def\mn@urlcharsother{\let\do\@makeother \do\$\do\&\do\#\do\^\do\_\do\%\do\~}
\def\mn@doi{\begingroup\mn@urlcharsother \@ifnextchar [ {\mn@doi@}
  {\mn@doi@[]}}
\def\mn@doi@[#1]#2{\def\@tempa{#1}\ifx\@tempa\@empty \href
  {http://dx.doi.org/#2} {doi:#2}\else \href {http://dx.doi.org/#2} {#1}\fi
  \endgroup}
\def\mn@eprint#1#2{\mn@eprint@#1:#2::\@nil}
\def\mn@eprint@arXiv#1{\href {http://arxiv.org/abs/#1} {{\tt arXiv:#1}}}
\def\mn@eprint@dblp#1{\href {http://dblp.uni-trier.de/rec/bibtex/#1.xml}
  {dblp:#1}}
\def\mn@eprint@#1:#2:#3:#4\@nil{\def\@tempa {#1}\def\@tempb {#2}\def\@tempc
  {#3}\ifx \@tempc \@empty \let \@tempc \@tempb \let \@tempb \@tempa \fi \ifx
  \@tempb \@empty \def\@tempb {arXiv}\fi \@ifundefined
  {mn@eprint@\@tempb}{\@tempb:\@tempc}{\expandafter \expandafter \csname
  mn@eprint@\@tempb\endcsname \expandafter{\@tempc}}}

\bibitem[\protect\citeauthoryear{Albert, van Weeren, Intema  \&
  R{\"{o}}ttgering}{Albert et~al.}{2020}]{Albert2020}
Albert J.~G.,  van Weeren R.~J.,  Intema H.~T.,   R{\"{o}}ttgering H. J.~A.,
  2020, \mn@doi [Astron. Astrophys.] {10.1051/0004-6361/201937424}, 635, A147

\bibitem[\protect\citeauthoryear{Ali et~al.,}{Ali et~al.}{2015}]{Ali2015}
Ali Z.~S.,  et~al., 2015, \mn@doi [Astrophys. J.] {10.1088/0004-637X/809/1/61},
  809, 61

\bibitem[\protect\citeauthoryear{Barry, Hazelton, Sullivan, Morales  \&
  Pober}{Barry et~al.}{2016}]{Barry2016}
Barry N.,  Hazelton B.,  Sullivan I.,  Morales M.~F.,   Pober J.~C.,  2016,
  \mn@doi [Mon. Not. R. Astron. Soc.] {10.1093/mnras/stw1380}, 461, 3135

\bibitem[\protect\citeauthoryear{Barry, Beardsley, Byrne, Hazelton, Morales,
  Pober  \& Sullivan}{Barry et~al.}{2019a}]{Barry2019a}
Barry N.,  Beardsley A.~P.,  Byrne R.,  Hazelton B.,  Morales M.~F.,  Pober
  J.~C.,   Sullivan I.,  2019a, \mn@doi [Publ. Astron. Soc. Aust.]
  {10.1017/pasa.2019.21}, 36, E026

\bibitem[\protect\citeauthoryear{Barry et~al.,}{Barry
  et~al.}{2019b}]{Barry2019}
Barry N.,  et~al., 2019b, \mn@doi [Astrophys. J.] {10.3847/1538-4357/ab40a8},
  884, 1

\bibitem[\protect\citeauthoryear{Beardsley et~al.,}{Beardsley
  et~al.}{2012}]{Beardsley2012}
Beardsley A.~P.,  et~al., 2012, \mn@doi [Mon. Not. R. Astron. Soc.]
  {10.1111/j.1365-2966.2012.20878.x}, 425, 1781

\bibitem[\protect\citeauthoryear{Berger et~al.,}{Berger
  et~al.}{2016}]{Berger2016}
Berger P.,  et~al., 2016, \mn@doi [Ground-based Airborne Telesc. VI]
  {10.1117/12.2233782}, 9906, 99060D

\bibitem[\protect\citeauthoryear{Byrne et~al.,}{Byrne et~al.}{2019}]{Byrne2019}
Byrne R.,  et~al., 2019, \mn@doi [Astrophys. J.] {10.3847/1538-4357/ab107d},
  875, 70

\bibitem[\protect\citeauthoryear{Datta, Bowman  \& Carilli}{Datta
  et~al.}{2010}]{Datta2010}
Datta A.,  Bowman J.~D.,   Carilli C.~L.,  2010, \mn@doi [Astrophys. J.]
  {10.1088/0004-637X/724/1/526}, 724, 526

\bibitem[\protect\citeauthoryear{DeBoer et~al.,}{DeBoer
  et~al.}{2017}]{DeBoer2017}
DeBoer D.~R.,  et~al., 2017, \mn@doi [Publ. Astron. Soc. Pacific]
  {10.1088/1538-3873/129/974/045001}, 129, 045001

\bibitem[\protect\citeauthoryear{Dillon \& Parsons}{Dillon \&
  Parsons}{2016}]{Dillon2016}
Dillon J.~S.,  Parsons A.~R.,  2016, \mn@doi [Astrophys. J.]
  {10.3847/0004-637X/826/2/181}, 826, 181

\bibitem[\protect\citeauthoryear{Dillon, Liu  \& Tegmark}{Dillon
  et~al.}{2013}]{Dillon2013}
Dillon J.~S.,  Liu A.,   Tegmark M.,  2013, \mn@doi [Phys. Rev. D]
  {10.1103/PhysRevD.87.043005}, 87, 043005

\bibitem[\protect\citeauthoryear{Dillon et~al.,}{Dillon
  et~al.}{2018}]{Dillon2018}
Dillon J.~S.,  et~al., 2018, \mn@doi [Mon. Not. R. Astron. Soc.]
  {10.1093/mnras/sty1060}, 477, 5670

\bibitem[\protect\citeauthoryear{Dillon et~al.,}{Dillon
  et~al.}{2020}]{Dillon2020}
Dillon J.~S.,  et~al., 2020, arXiv e-prints

\bibitem[\protect\citeauthoryear{Ewall-Wice, Dillon, Liu  \& Hewitt}{Ewall-Wice
  et~al.}{2016}]{Ewall-Wice2017}
Ewall-Wice A.,  Dillon J.~S.,  Liu A.,   Hewitt J.,  2016, \mn@doi [Mon. Not.
  R. Astron. Soc.] {10.1093/mnras/stx1221}, 470, 1849

\bibitem[\protect\citeauthoryear{Furlanetto, {Peng Oh}  \& Briggs}{Furlanetto
  et~al.}{2006}]{Furlanetto2006}
Furlanetto S.~R.,  {Peng Oh} S.,   Briggs F.~H.,  2006, \mn@doi [Phys. Rep.]
  {10.1016/j.physrep.2006.08.002}, 433, 181

\bibitem[\protect\citeauthoryear{Grobler, Nunhokee, Smirnov, van Zyl  \& de
  Bruyn}{Grobler et~al.}{2014}]{Grobler2014}
Grobler T.~L.,  Nunhokee C.~D.,  Smirnov O.~M.,  van Zyl A.~J.,   de Bruyn
  A.~G.,  2014, \mn@doi [Mon. Not. R. Astron. Soc.] {10.1093/mnras/stu268},
  439, 4030

\bibitem[\protect\citeauthoryear{Grobler, Stewart, Wijnholds, Kenyon  \&
  Smirnov}{Grobler et~al.}{2016}]{Grobler2016}
Grobler T.~L.,  Stewart A.~J.,  Wijnholds S.~J.,  Kenyon J.~S.,   Smirnov
  O.~M.,  2016, \mn@doi [Mon. Not. R. Astron. Soc.] {10.1093/mnras/stw1437},
  461, 2975

\bibitem[\protect\citeauthoryear{Grobler, Bernardi, Kenyon, Parsons  \&
  Smirnov}{Grobler et~al.}{2018}]{Grobler2018}
Grobler T.~L.,  Bernardi G.,  Kenyon J.~S.,  Parsons A.~R.,   Smirnov O.~M.,
  2018, \mn@doi [Mon. Not. R. Astron. Soc.] {10.1093/mnras/sty357}, 476, 2410

\bibitem[\protect\citeauthoryear{Hamaker}{Hamaker}{2000}]{Hamaker2000}
Hamaker J.~P.,  2000, \mn@doi [Astron. Astrophys. Suppl. Ser.]
  {10.1051/aas:2000337}, 143, 515

\bibitem[\protect\citeauthoryear{Hamaker, Bregman  \& Sault}{Hamaker
  et~al.}{1996}]{Hamaker1996a}
Hamaker J.~P.,  Bregman J.~D.,   Sault R.~J.,  1996, \mn@doi [Astron.
  Astrophys. Suppl. Ser.] {10.1051/aas:1996146}, 117, 137

\bibitem[\protect\citeauthoryear{Hazelton, Morales  \& Sullivan}{Hazelton
  et~al.}{2013}]{Hazelton2013}
Hazelton B.~J.,  Morales M.~F.,   Sullivan I.~S.,  2013, \mn@doi [Astrophys.
  J.] {10.1088/0004-637X/770/2/156}, 770, 156

\bibitem[\protect\citeauthoryear{Hurley-Walker et~al.,}{Hurley-Walker
  et~al.}{2017}]{hurley-walker2017}
Hurley-Walker N.,  et~al., 2017, \mn@doi [Mon. Not. R. Astron. Soc.]
  {10.1093/mnras/stw2337}, 464, 1146

\bibitem[\protect\citeauthoryear{Jagannathan, Bhatnagar, Rau  \&
  Taylor}{Jagannathan et~al.}{2017}]{Jagannathan2017}
Jagannathan P.,  Bhatnagar S.,  Rau U.,   Taylor A.~R.,  2017, \mn@doi [Astron.
  J.] {10.3847/1538-3881/aa77f8}, 154, 56

\bibitem[\protect\citeauthoryear{Jordan et~al.,}{Jordan
  et~al.}{2017}]{Jordan2017}
Jordan C.~H.,  et~al., 2017, \mn@doi [Mon. Not. R. Astron. Soc.]
  {10.1093/mnras/stx1797}, 471, 3974

\bibitem[\protect\citeauthoryear{Joseph, Trott  \& Wayth}{Joseph
  et~al.}{2018}]{Joseph2018}
Joseph R.~C.,  Trott C.~M.,   Wayth R.~B.,  2018, \mn@doi [Astron. J.]
  {10.3847/1538-3881/aaec0b}, 156, 285

\bibitem[\protect\citeauthoryear{Joseph, Trott, Wayth  \& Nasirudin}{Joseph
  et~al.}{2020}]{Joseph2020}
Joseph R.~C.,  Trott C.~M.,  Wayth R.~B.,   Nasirudin A.,  2020, \mn@doi [Mon.
  Not. R. Astron. Soc.] {10.1093/mnras/stz3375}, 492, 2017

\bibitem[\protect\citeauthoryear{Kazemi \& Yatawatta}{Kazemi \&
  Yatawatta}{2013}]{Kazemi2013}
Kazemi S.,  Yatawatta S.,  2013, \mn@doi [Mon. Not. R. Astron. Soc.]
  {10.1093/mnras/stt1347}, 435, 597

\bibitem[\protect\citeauthoryear{Kazemi, Yatawatta, Zaroubi, Lampropoulos, de
  Bruyn, Koopmans  \& Noordam}{Kazemi et~al.}{2011}]{Kazemi2011}
Kazemi S.,  Yatawatta S.,  Zaroubi S.,  Lampropoulos P.,  de Bruyn A.~G.,
  Koopmans L. V.~E.,   Noordam J.,  2011, \mn@doi [Mon. Not. R. Astron. Soc.]
  {10.1111/j.1365-2966.2011.18506.x}, 414, 1656

\bibitem[\protect\citeauthoryear{Kazemi, Yatawatta  \& Zaroubi}{Kazemi
  et~al.}{2013}]{Kazemi2013a}
Kazemi S.,  Yatawatta S.,   Zaroubi S.,  2013, \mn@doi [Mon. Not. R. Astron.
  Soc.] {10.1093/mnras/stt018}, 430, 1457

\bibitem[\protect\citeauthoryear{Kern et~al.,}{Kern et~al.}{2020}]{Kern2019}
Kern N.~S.,  et~al., 2020, \mn@doi [Astrophys. J.] {10.3847/1538-4357/ab67bc},
  890, 122

\bibitem[\protect\citeauthoryear{Kohn et~al.,}{Kohn et~al.}{2019}]{Kohn2019}
Kohn S.~A.,  et~al., 2019, \mn@doi [Astrophys. J.] {10.3847/1538-4357/ab2f72},
  882, 58

\bibitem[\protect\citeauthoryear{Lanman, Hazelton, Jacobs, Kolopanis, Pober,
  Aguirre  \& Thyagarajan}{Lanman et~al.}{2019}]{Lanman2019}
Lanman A.,  Hazelton B.,  Jacobs D.,  Kolopanis M.,  Pober J.,  Aguirre J.,
  Thyagarajan N.,  2019, \mn@doi [J. Open Source Softw.] {10.21105/joss.01234},
  4, 1234

\bibitem[\protect\citeauthoryear{Li et~al.,}{Li et~al.}{2018}]{Li2018}
Li W.,  et~al., 2018, \mn@doi [Astrophys. J.] {10.3847/1538-4357/aad3c3}, 863,
  170

\bibitem[\protect\citeauthoryear{Li et~al.,}{Li et~al.}{2019}]{Li2019}
Li W.,  et~al., 2019, \mn@doi [Astrophys. J.] {10.3847/1538-4357/ab55e4}, 887,
  141

\bibitem[\protect\citeauthoryear{Liu, Tegmark, Morrison, Lutomirski  \&
  Zaldarriaga}{Liu et~al.}{2010}]{Liu2010}
Liu A.,  Tegmark M.,  Morrison S.,  Lutomirski A.,   Zaldarriaga M.,  2010,
  \mn@doi [Mon. Not. R. Astron. Soc.] {10.1111/j.1365-2966.2010.17174.x}, 408,
  1029

\bibitem[\protect\citeauthoryear{Mertens et~al.,}{Mertens
  et~al.}{2020}]{Mertens2020}
Mertens F.~G.,  et~al., 2020, \mn@doi [Mon. Not. R. Astron. Soc.]
  {10.1093/mnras/staa327}, 493, 1662

\bibitem[\protect\citeauthoryear{Mitchell, Greenhill, Wayth, Sault, Lonsdale,
  Cappallo, Morales  \& Ord}{Mitchell et~al.}{2008}]{Mitchell2008}
Mitchell D.,  Greenhill L.,  Wayth R.,  Sault R.,  Lonsdale C.,  Cappallo R.,
  Morales M.,   Ord S.,  2008, \mn@doi [IEEE J. Sel. Top. Signal Process.]
  {10.1109/JSTSP.2008.2005327}, 2, 707

\bibitem[\protect\citeauthoryear{Morales \& Wyithe}{Morales \&
  Wyithe}{2010}]{Morales2010}
Morales M.~F.,  Wyithe J. S.~B.,  2010, \mn@doi [Annu. Rev. Astron. Astrophys.]
  {10.1146/annurev-astro-081309-130936}, 48, 127

\bibitem[\protect\citeauthoryear{Morales, Hazelton, Sullivan  \&
  Beardsley}{Morales et~al.}{2012}]{Morales2012}
Morales M.~F.,  Hazelton B.,  Sullivan I.,   Beardsley A.,  2012, \mn@doi
  [Astrophys. J.] {10.1088/0004-637X/752/2/137}, 752

\bibitem[\protect\citeauthoryear{Mort, Dulwich, Salvini, Adami  \& Jones}{Mort
  et~al.}{2010}]{Mort2010}
Mort B.~J.,  Dulwich F.,  Salvini S.,  Adami K.~Z.,   Jones M.~E.,  2010, in
  IEEE Int. Symp.. pp 690--694, \mn@doi{10.1109/ARRAY.2010.5613289}

\bibitem[\protect\citeauthoryear{Newburgh et~al.,}{Newburgh
  et~al.}{2014}]{Newburgh2014}
Newburgh L.~B.,  et~al., 2014, \mn@doi [Ground-based Airborne Telesc. V]
  {10.1117/12.2056962}, 9145, 91454V

\bibitem[\protect\citeauthoryear{Newburgh et~al.,}{Newburgh
  et~al.}{2016}]{Newburgh2016}
Newburgh L.~B.,  et~al., 2016, \mn@doi [Ground-based Airborne Telesc. VI]
  {10.1117/12.2234286}, 9906, 99065X

\bibitem[\protect\citeauthoryear{Ollier, {El Korso}, Boyer, Larzabal  \&
  Pesavento}{Ollier et~al.}{2017}]{Ollier2017}
Ollier V.,  {El Korso} M.~N.,  Boyer R.,  Larzabal P.,   Pesavento M.,  2017,
  \mn@doi [IEEE Trans. Signal Process.] {10.1109/TSP.2017.2733496}, 65, 5649

\bibitem[\protect\citeauthoryear{Orosz, Dillon, Ewall-Wice, Parsons  \&
  Thyagarajan}{Orosz et~al.}{2019}]{Orosz2019}
Orosz N.,  Dillon J.~S.,  Ewall-Wice A.,  Parsons A.~R.,   Thyagarajan N.,
  2019, \mn@doi [Mon. Not. R. Astron. Soc.] {10.1093/mnras/stz1287}, 487, 537

\bibitem[\protect\citeauthoryear{Parsons et~al.,}{Parsons
  et~al.}{2010}]{Parsons2010}
Parsons A.~R.,  et~al., 2010, \mn@doi [Astron. J.]
  {10.1088/0004-6256/139/4/1468}, 139, 1468

\bibitem[\protect\citeauthoryear{Parsons, Pober, Aguirre, Carilli, Jacobs  \&
  Moore}{Parsons et~al.}{2012}]{Parsons2012}
Parsons A.~R.,  Pober J.~C.,  Aguirre J.~E.,  Carilli C.~L.,  Jacobs D.~C.,
  Moore D.~F.,  2012, \mn@doi [Astrophys. J.] {10.1088/0004-637X/756/2/165},
  756

\bibitem[\protect\citeauthoryear{Patil et~al.,}{Patil et~al.}{2016}]{Patil2016}
Patil A.~H.,  et~al., 2016, \mn@doi [Mon. Not. R. Astron. Soc.]
  {10.1093/mnras/stw2277}, 463, 4317

\bibitem[\protect\citeauthoryear{Pen, Chang, Hirata, Peterson, Roy, Gupta,
  Odegova  \& Sigurdson}{Pen et~al.}{2009}]{Pen2009}
Pen U.~L.,  Chang T.~C.,  Hirata C.~M.,  Peterson J.~B.,  Roy J.,  Gupta Y.,
  Odegova J.,   Sigurdson K.,  2009, \mn@doi [Mon. Not. R. Astron. Soc.]
  {10.1111/j.1365-2966.2009.14980.x}, 399, 181

\bibitem[\protect\citeauthoryear{Salvini \& Wijnholds}{Salvini \&
  Wijnholds}{2014}]{Salvini2014}
Salvini S.,  Wijnholds S.~J.,  2014, in 2014 31th URSI Gen. Assem. Sci. Symp.
  URSI GASS 2014. , \mn@doi{10.1109/URSIGASS.2014.6930038}

\bibitem[\protect\citeauthoryear{Sault, Hamaker  \& Bregman}{Sault
  et~al.}{1996}]{Sault1996}
Sault R.~J.,  Hamaker J.~P.,   Bregman J.~D.,  1996, \mn@doi [Astron.
  Astrophys. Suppl. Ser.] {10.1051/aas:1996100}, 117, 149

\bibitem[\protect\citeauthoryear{Sievers}{Sievers}{2017}]{Sievers2017}
Sievers J.~L.,  2017, arXiv e-prints

\bibitem[\protect\citeauthoryear{Sob, Bester, Smirnov, Kenyon  \& Grobler}{Sob
  et~al.}{2020}]{Sob2020}
Sob U.~M.,  Bester H.~L.,  Smirnov O.~M.,  Kenyon J.~S.,   Grobler T.~L.,
  2020, \mn@doi [Mon. Not. R. Astron. Soc.] {10.1093/mnras/stz3037}, 491, 1026

\bibitem[\protect\citeauthoryear{Sullivan et~al.,}{Sullivan
  et~al.}{2012}]{Sullivan2012}
Sullivan I.~S.,  et~al., 2012, \mn@doi [Astrophys. J.]
  {10.1088/0004-637X/759/1/17}, 759

\bibitem[\protect\citeauthoryear{Tasse et~al.,}{Tasse et~al.}{2018}]{Tasse2018}
Tasse C.,  et~al., 2018, \mn@doi [Astron. Astrophys.]
  {10.1051/0004-6361/201731474}, 611

\bibitem[\protect\citeauthoryear{Tegmark, Hamilton, Strauss, Vogeley  \&
  Szalay}{Tegmark et~al.}{1998}]{Tegmark1998}
Tegmark M.,  Hamilton A. J.~S.,  Strauss M.~A.,  Vogeley M.~S.,   Szalay A.~S.,
   1998, \mn@doi [Astrophys. J.] {10.1086/305663}, 499, 555

\bibitem[\protect\citeauthoryear{Thyagarajan et~al.,}{Thyagarajan
  et~al.}{2013}]{Thyagarajan2013}
Thyagarajan N.,  et~al., 2013, \mn@doi [Astrophys. J.]
  {10.1088/0004-637X/776/1/6}, 776

\bibitem[\protect\citeauthoryear{Thyagarajan et~al.,}{Thyagarajan
  et~al.}{2015}]{Thyagarajan2015}
Thyagarajan N.,  et~al., 2015, \mn@doi [Astrophys. J.]
  {10.1088/0004-637X/804/1/14}, 804

\bibitem[\protect\citeauthoryear{Tingay et~al.,}{Tingay
  et~al.}{2013}]{Tingay2013}
Tingay S.~J.,  et~al., 2013, \mn@doi [Publ. Astron. Soc. Aust.]
  {10.1017/pasa.2012.007}, 30

\bibitem[\protect\citeauthoryear{Trott, Wayth  \& Tingay}{Trott
  et~al.}{2012}]{Trott2012}
Trott C.~M.,  Wayth R.~B.,   Tingay S.~J.,  2012, \mn@doi [Astrophys. J.]
  {10.1088/0004-637X/757/1/101}, 757

\bibitem[\protect\citeauthoryear{Vanderlinde et~al.,}{Vanderlinde
  et~al.}{2019}]{Vanderlinde2019}
Vanderlinde K.,  et~al., 2019, \mn@doi [arXiv] {10.5281/zenodo.3765414}

\bibitem[\protect\citeauthoryear{Vedantham, {Udaya Shankar}  \&
  Subrahmanyan}{Vedantham et~al.}{2012}]{Vedantham2012}
Vedantham H.,  {Udaya Shankar} N.,   Subrahmanyan R.,  2012, \mn@doi
  [Astrophys. J.] {10.1088/0004-637X/745/2/176}, 745

\bibitem[\protect\citeauthoryear{Wayth et~al.,}{Wayth et~al.}{2018}]{Wayth2018}
Wayth R.~B.,  et~al., 2018, \mn@doi [Publ. Astron. Soc. Aust.]
  {10.1017/pasa.2018.37}

\bibitem[\protect\citeauthoryear{Wieringa}{Wieringa}{1992}]{Wieringa1992}
Wieringa M.~H.,  1992, \mn@doi [Exp. Astron.] {10.1007/BF00420576}, 2, 203

\bibitem[\protect\citeauthoryear{Wijnholds, Grobler  \& Smirnov}{Wijnholds
  et~al.}{2016}]{Wijnholds2016}
Wijnholds S.~J.,  Grobler T.~L.,   Smirnov O.~M.,  2016, \mn@doi [Mon. Not. R.
  Astron. Soc.] {10.1093/mnras/stw118}, 457, 2331

\bibitem[\protect\citeauthoryear{Yatawatta}{Yatawatta}{2015}]{Yatawatta2015}
Yatawatta S.,  2015, \mn@doi [Mon. Not. R. Astron. Soc.]
  {10.1093/mnras/stv596}, 449, 4506

\bibitem[\protect\citeauthoryear{Zhang et~al.,}{Zhang et~al.}{2020}]{Zhang2020}
Zhang Z.,  et~al., 2020, \mn@doi [Publications of the Astronomical Society of
  Australia] {10.1017/pasa.2020.37}, 37, e045

\bibitem[\protect\citeauthoryear{Zheng et~al.,}{Zheng et~al.}{2014}]{Zheng2014}
Zheng H.,  et~al., 2014, \mn@doi [Mon. Not. R. Astron. Soc.]
  {10.1093/mnras/stu1773}, 445, 1084

\bibitem[\protect\citeauthoryear{van Weeren et~al.,}{van Weeren
  et~al.}{2016}]{Weeren2016}
van Weeren R.~J.,  et~al., 2016, \mn@doi [Astrophys. J. Suppl. Ser.]
  {10.3847/0067-0049/223/1/2}, 223, 2

\makeatother
\end{thebibliography}

\appendix

\section{Working With Singular Covariance Matrices}
\label{app:perfect_redundancy}

In this section we explain how to use Singular Value Decomposition (SVD) to resolve the problem of non-invertible covariance matrices by remapping the calibration parameters to a well-defined basis. This approach is fully general and can be applied to any singular covariance matrix, not just those corresponding to traditional redundant calibration as described in \S\ref{s:red_cal}. It remaps an initial calibration parameter basis $\fitparamsu_\text{orig}$, where the associated covariance matrix $\modelcov{}_\text{,orig} = \operatorname{cov}[\fitparamsu_\text{orig} \fitparamsu_\text{orig}^\dag]$ is singular, to a new basis $\fitparamsu$ with an invertible covariance matrix $\modelcov = \operatorname{cov}[\fitparamsu \fitparamsu^\dag]$.

All covariance matrices have eigenvalues $\lambda_j \ge 0$ and eigenvectors that span the space. We can therefore transform $\fitparamsu_\text{orig}$ into the basis of eigenvectors of $\modelcov{}_\text{,orig}$. We define a new vector $\fitparamsu_\text{diag}$ as $\fitparamsu_\text{orig}$ expressed in the orthogonal eigenbasis of $\modelcov{}_\text{,orig}$:
\begin{equation}
    \fitparamsu_\text{diag} = \boldsymbol{\mathsf{E}} \fitparamsu_\text{orig}
\end{equation}
where
\begin{equation}
    \boldsymbol{\mathsf{E}} = \begin{bmatrix}
    \boldsymbol{e}_1^\dag \\ 
    \boldsymbol{e}_2^\dag \\ 
    \vdots \\ \boldsymbol{e}_{N_\text{orig}}^\dag
    \end{bmatrix}
\end{equation}
and $\boldsymbol{e}_j$ are the orthogonal eigenvectors of $\modelcov{}_\text{,orig}$.
Here $N_\text{orig}$ is the length of $\fitparamsu_\text{orig}$. The associated covariance matrix is
\begin{equation}
    \modelcov{}_\text{,diag} = \operatorname{cov} [\fitparamsu_\text{diag},  \fitparamsu_\text{diag}^\dag] = \boldsymbol{\mathsf{E}} \modelcov{}_{,\text{orig}} \boldsymbol{\mathsf{E}}^{-1} = \operatorname{diag}(\begin{bmatrix}
    \lambda_1 & \lambda_2 & \cdots &  \lambda_{N_\text{orig}}
    \end{bmatrix}),
\end{equation}
where $\lambda_j$ are the eigenvalues of $\modelcov{}_\text{,orig}$.

If $\modelcov{}_{,\text{orig}}$ is singular than it will have $\lambda_j = 0$ for some $j$. This has physical meaning. If $\lambda_j = 0$ then $\operatorname{var}[\fitparamsu_{\text{diag},j}] = 0$. It follows that those elements of $\fitparamsu_{\text{diag}}$ must take their expectation values: $\fitparamsu_{\text{diag},j} = \langle \fitparamsu_{\text{diag},j} \rangle$. We assume that $\langle \fitparamsu_\text{orig} \rangle = \modelvals_\text{orig}$, so likewise $\langle \fitparamsu_\text{diag} \rangle = \modelvals_\text{diag}$ where
\begin{equation}
    \modelvals_\text{diag} = \boldsymbol{\mathsf{E}} \modelvals_\text{orig}.
\end{equation}
Therefore $\fitparamsu_{\text{diag},j} = \modelvals_{\text{diag},j}$ for all $j$ where $\lambda_j = 0$. Those calibration parameters simply take on their modeled values; they are no longer fit during the calibration process.

We can now restrict the basis of the calibration parameters to only those that can vary in calibration. The number of independent calibration parameters is $N_\text{red}$, which is equal to the rank of $\modelcov{}_\text{,orig}$. The new basis of calibration parameters is given by
\begin{equation}
    \fitparamsu = \boldsymbol{\mathsf{T}} \fitparamsu_\text{orig}.
\label{eq:svd_fitparams}
\end{equation}
Here $\boldsymbol{\mathsf{T}}$ is
\begin{equation}
    \boldsymbol{\mathsf{T}} = \begin{bmatrix}
    \boldsymbol{e}_1^\dag \\ 
    \boldsymbol{e}_2^\dag \\ 
    \vdots \\ \boldsymbol{e}_{N_\text{red}}^\dag
    \end{bmatrix},
\end{equation}
where we use only $\boldsymbol{e}_j$ for $\lambda_j \neq 0$. $\modelcov = \operatorname{cov}[\fitparamsu, \fitparamsu^\dag]$ is now diagonal and invertible.

To develop physical intuition for this remapping we can consider the case of a perfectly redundant array as described in \S\ref{s:red_cal}, where $\fitparamsu_\text{orig}$ correspond to the full set of visibilities. Here $\operatorname{cov}[\fitparamsu_{\text{orig},j}, \fitparamsu_{\text{orig},k}^*] = \operatorname{var}[\fitparamsu_{\text{orig},j}]$ if baselines $j$ and $k$ belong to the same redundant set and $\operatorname{cov}[\fitparamsu_{\text{orig},j}, \fitparamsu_{\text{orig},k}^*] = 0$ otherwise. In this case, eigenvectors $\boldsymbol{e}_j$ for $\lambda_j \neq 0$ represent the average visibilities from redundant baseline sets. Eigenvectors $\boldsymbol{e}_j$ for $\lambda_j = 0$ represent visibility differences within redundant baseline sets.

$\boldsymbol{\mathsf{T}}$ is a rectangular matrix with linearly independent rows. This means that a right pseudoinverse $\boldsymbol{\mathsf{T}}^\text{p}$ exists such that $\boldsymbol{\mathsf{T}} \boldsymbol{\mathsf{T}}^\text{p} = \mathbb{1}$. However, it does not follow that $\boldsymbol{\mathsf{T}}^\text{p} \boldsymbol{\mathsf{T}} = \mathbb{1}$, so inverting Equation \ref{eq:svd_fitparams} is not straightforward. Rather, we get the somewhat unwieldy expression
\begin{equation}
    \fitparamsu_\text{orig} = \boldsymbol{\mathsf{T}}^\text{p} \fitparamsu + \sum_{\substack{\text{$j$ for all} \\ \lambda_{j}=0}} \modelvals_{\text{diag},j} \frac{\boldsymbol{e}_j}{|\boldsymbol{e}_j|^2}.
\end{equation}
This expression simplifies when we assume that $\modelvals_{\text{diag},j} = 0$ when $\lambda_{j}=0$ --- in other words, that the model values $\modelvals_\text{orig}$ are orthogonal to the null space of $\modelcov{}_\text{,orig}$. We then get that
\begin{equation}
    \fitparamsu_\text{orig} = \boldsymbol{\mathsf{T}}^\text{p} \fitparamsu.
\end{equation}
This is a physically-motivated assumption. For the case of a perfectly redundant array, requiring that $\modelvals_\text{orig}$ is orthogonal to the null space means requiring that model visibilities from different baselines within redundant baseline sets are equal. Components of $\modelvals_\text{orig}$ in the null space indicate a disagreement between the instrument models used to produce $\modelvals_\text{orig}$ and $\modelcov{}_\text{,orig}$.

\section{Exploring Gain Amplitude Biases}
\label{app:gain_amps}

In \S\ref{s:sim_results} we demonstrate that, in simulation, traditional sky-based and redundant calibration methods fit gain solutions $\hat{\gains}$ that are, on average, biased low in their amplitudes. Unified calibration solutions do not exhibit this effect. In this appendix we explore the gain amplitude bias using analytic techniques. We show that suppression of the average gain amplitude in redundant and sky-based calibration stems from decoherence between the measured visibilities and the model visibilities in the presence of sky model error.

\subsection{Sky-Based Calibration}
\label{s:sky_cal_gain_amp}

Following Equation \ref{eq:sky_cal}, traditional sky-based calibration consists of minimizing the negative log-likelihood
\begin{equation}
    \negloglikelihood(\gains) =  \frac{1}{\sigma_\text{T}^2} \sum_\alpha \sum_{\{b,c\} \in \alpha} \left|\data_{bc} - \gains_b \gains_c^*  \modelvals_\alpha \right|^2.
\end{equation}
Here $b$ and $c$ index antennas and $\alpha$ indexed redundant baseline sets. $\sum_{\{b,c\} \in \alpha}$ indicates the sum over all baselines $\{b,c\}$ within redundant set $\alpha$. For a non-regular array, each `redundant baseline set' will contain just one baseline. We assume no baseline-dependent weighting, such that $\sigma_\text{T}$ is a constant.

We can derive an analytic expression for the maximum likelihood solution $\hat{g}$ by imposing the extremum condition
\begin{equation}
    \left. \frac{\partial \negloglikelihood(\gains)}{\partial \gains_c} \right|_{\gains=\hat{\gains}} = 0.
\end{equation}
Differentiating the negative log-likelihood with respect to the complex gains and evaluating gives
\begin{equation}
    \hat{\gains}_c = \frac{\sum_\alpha \sum_{b\text{ for } {\{b,c\} \in \alpha}} \data_{bc}^* \hat{\gains}_b \modelvals_\alpha}{\sum_\alpha \sum_{b\text{ for } {\{b,c\} \in \alpha}} |\hat{\gains}_b \modelvals_\alpha|^2},
\label{eq:sky_cal_gains}
\end{equation}
where $\sum_{b\text{ for } {\{b,c\} \in \alpha}}$ denotes choosing antenna $b$ such that baseline $\{b,c\}$ belongs to redundant set $\alpha$.
Equation \ref{eq:sky_cal_gains} represents a nonlinear system of $N_\text{ant}$ coupled equations. Solving these equations together gives the calibration solutions. However, as this is not analytically tractable, practical calibration algorithms employ numerical optimization techniques.

We define a new variable $\gainamp$ equal to the average gain amplitude
\begin{equation}
    \gainamp = \frac{1}{N_\text{ant}} \sum_c |\gains_c|
\end{equation}
and parameterize the gains as $\gains = \gainamp \unnormgains$. Plugging this into equation \ref{eq:sky_cal_gains}, we get
\begin{equation}
    \hat{\gainamp}^2 = \frac{1}{N_\text{ant}} \sum_c \frac{\left| \sum_\alpha \sum_{b\text{ for } {\{b,c\} \in \alpha}} \data_{bc}^* \hat{\unnormgains}_b \modelvals_\alpha \right|}{\sum_\alpha \sum_{b\text{ for } {\{b,c\} \in \alpha}} |\hat{\unnormgains}_b \modelvals_\alpha|^2}.
\end{equation}

At this point we make a series of simplifying approximations. As in the simulations in \S\ref{s:sim_results}, we allow that the true gains are equal to 1. We can therefore make a first-order approximation of $\hat{\gainamp}$ by setting $\hat{\unnormgains} = 1$:
\begin{equation}
    \hat{\gainamp}^2 \approx \frac{1}{N_\text{ant}} \sum_c \frac{\left| \sum_\alpha \sum_{b\text{ for } {\{b,c\} \in \alpha}} \data_{bc}^* \modelvals_\alpha \right|}{\sum_\alpha |\modelvals_\alpha|^2}.
\end{equation}
This can be rewritten as
\begin{equation}
    \hat{\gainamp}^2 \approx \frac{\left| \langle \data^* \modelvals \rangle \right|}{\langle |\modelvals|^2 \rangle},
\end{equation}
where $\langle \rangle$ denotes the average across redundant baseline sets.

We describe the model visibilities $\modelvals$ as a `true' visibility $\fitparamsu_\text{true}$ with error $\boldsymbol{n}_\text{M}$:
\begin{equation}
    \modelvals_\alpha = \fitparamsu_{\text{true,}\alpha} + \boldsymbol{n}_{\text{M,}\alpha}.
\label{eq:model_decomp}
\end{equation}
Likewise, we represent the data as
\begin{equation}
    \data_{bc} = \fitparamsu_{\text{true,}\alpha} + \boldsymbol{n}_{\text{T,}bc}
\label{eq:data_decomp}
\end{equation}
where $\boldsymbol{n}_\text{T}$ represents the thermal noise and baseline $\{b,c\}$ belongs to redundant set $\alpha$. If we assume that $\boldsymbol{n}_\text{M}$, $\boldsymbol{n}_\text{T}$, and $\fitparamsu_\text{true}$ are uncorrelated with one another, then the average quantities
\begin{equation}
    \langle \data^* \modelvals \rangle = \langle |\fitparamsu_\text{true}|^2 \rangle
\end{equation}
and
\begin{equation}
    \langle |\modelvals|^2 \rangle = \langle |\fitparamsu_\text{true}|^2 \rangle + \langle |\boldsymbol{n}_\text{M}|^2 \rangle.
\end{equation}
We thereby estimate that
\begin{equation}
    \hat{\gainamp} \approx \sqrt{ \frac{\langle |\fitparamsu_\text{true}|^2 \rangle}{\langle |\fitparamsu_\text{true}|^2 \rangle + \langle |\boldsymbol{n}_\text{M}|^2 \rangle}} \le 1.
\label{eq:sky_cal_gain_amp_approx}
\end{equation}
Decoherence of the sky model error and thermal noise means that we expect suppression of average amplitude of the fit gains.

We can compare this approximation to the results of the simulation in \S\ref{s:random_error_sim}. Across all trials in that simulation we calculate that $\left\langle |\fitparamsu_\text{true}|^2 \right\rangle = 38.45 \text{ Jy}^2$ and $\left\langle |\boldsymbol{n}_\text{M}|^2 \right\rangle = 0.31 \text{ Jy}^2$. Plugging these into Equation \ref{eq:sky_cal_gain_amp_approx}, we estimate $\gainamp \approx 0.9960$. The average gain amplitude from the sky-based calibration simulation is $0.9964$.

\subsection{Redundant Calibration}

The overall amplitude of the gains is degenerate in the relative calibration step of redundant calibration (Equation \ref{eq:red_cal}). The average gain amplitude is therefore fit in absolute calibration by minimizing the negative log-likelihood
\begin{equation}
    \negloglikelihood(\gainamp) = \frac{1}{\sigma_\text{T}^2} \sum_\alpha \sum_{\{b, c\} \in \alpha} \left|\data_{bc} - \gainamp^2 \hat{\unnormgains}_b \hat{\unnormgains}_c^* \modelvals_\alpha \right|^2,
\label{eq:abs_cal_amp}
\end{equation}
where $\hat{\unnormgains}$ have been calculated from relative calibration. For simplicity, we also assume that the gains' phase gradient across the array has been constrained in a separate absolute calibration step. Applying the extremum condition
\begin{equation}
    \left. \frac{\partial \negloglikelihood(\gainamp)}{\partial \gainamp} \right|_{\gainamp=\hat{\gainamp}} = 0,
\end{equation}
we find that
\begin{equation}
    \hat{\gainamp}^2 = \frac{\sum_\alpha \sum_{\{b, c\} \in \alpha} \operatorname{Re}(\data_{bc}^* \hat{\unnormgains}_b \hat{\unnormgains}_c^* \modelvals_{\alpha})}{ \sum_\alpha \sum_{\{b, c\} \in \alpha} |\hat{\unnormgains}_b \hat{\unnormgains}_c^* \modelvals_{\alpha}|^2}.
\end{equation}

As in \S\ref{s:sky_cal_gain_amp}, we take the first-order approximation that $\hat{\unnormgains} = 1$, giving
\begin{equation}
    \hat{\gainamp}^2 \approx \frac{\sum_\alpha \sum_{\{b, c\} \in \alpha} \operatorname{Re}(\data_{bc}^* \modelvals_{\alpha})}{ \sum_\alpha \sum_{\{b, c\} \in \alpha} |\modelvals_{\alpha}|^2} \approx \frac{\langle \operatorname{Re}(\data^* \modelvals) \rangle}{\langle |\modelvals|^2 \rangle}.
\end{equation}
Using the decompositions in Equations \ref{eq:model_decomp} and \ref{eq:data_decomp}, and once again assuming no correlation between $\boldsymbol{n}_\text{M}$, $\boldsymbol{n}_\text{T}$, and $\fitparamsu_\text{true}$, we get a result identical to Equation \ref{eq:sky_cal_gain_amp_approx}:
\begin{equation}
    \hat{\gainamp} \approx \sqrt{ \frac{\langle |\fitparamsu_\text{true}|^2 \rangle}{\langle |\fitparamsu_\text{true}|^2 \rangle + \langle |\boldsymbol{n}_\text{M}|^2 \rangle}} \le 1.
\end{equation}
We expect redundant calibration to return approximately the same average gain amplitude as sky-based calibration. Indeed, in the simulation presented in \S\ref{s:random_error_sim}, the average gain amplitude is $0.9964$ for each sky-based and redundant calibration.

We note that Equation \ref{eq:abs_cal_amp} represents a commonly-used absolute calibration implementation, but it is not the only method of constraining redundant calibration's gain amplitude. Future work could explore alternative approaches to absolute calibration aimed at mitigating gain amplitude bias.

\subsection{Unified Calibration}

From Equation \ref{eq:cal_sim}, the unified calibration framework explored in simulation in \S\ref{s:sim_results} consists of minimizing the negative log-likelihood
\begin{equation}
    \negloglikelihood(\gains,\fitparamsu) =  \frac{1}{\sigma_\text{T}^2} \sum_\alpha \sum_{\{b,c\} \in \alpha} \left|\data_{bc} - \gains_b \gains_c^* \fitparamsu_\alpha \right|^2 + \frac{1}{\sigma_\text{M}^2} \sum_\alpha \sum_\gamma (\modelcorr^{-1})_{\alpha\gamma} (\fitparamsu_\alpha - \modelvals_\alpha) (\fitparamsu_\gamma^* - \modelvals_\gamma^*),
\end{equation}
where now $\alpha$ and $\gamma$ index redundant baseline sets. For the purposes of this discussion we will approximate $\modelcorr \approx \mathbb{1}$, neglecting the covariance between redundant baseline sets. The negative log-likelihood then simplifies to
\begin{equation}
    \negloglikelihood(\gains,\fitparamsu) =  \frac{1}{\sigma_\text{T}^2} \sum_\alpha \sum_{\{b,c\} \in \alpha} \left|\data_{bc} - \gains_b \gains_c^* \fitparamsu_\alpha \right|^2 + \frac{1}{\sigma_\text{M}^2} \sum_\alpha |\fitparamsu_\alpha - \modelvals_\alpha|^2.
\end{equation}

Solving the extremum condition
\begin{equation}
    \left. \frac{\partial \negloglikelihood(\gains, \fitparamsu)}{\partial \gains_c} \right|_{\gains=\hat{\gains}, \fitparamsu=\hat{\fitparamsu}} = 0.
\end{equation}
gives
\begin{equation}
    \hat{\gains}_c = \frac{\sum_\alpha \sum_{b\text{ for } {\{b,c\} \in \alpha}} \data_{bc}^* \hat{\gains}_b \hat{\fitparamsu}_\alpha}{\sum_\alpha \sum_{b\text{ for } {\{b,c\} \in \alpha}} |\hat{\gains}_b \hat{\fitparamsu}_\alpha|^2},
\end{equation}
from which it follows that
\begin{equation}
    \hat{\gainamp} = \frac{1}{N_\text{ant}} \sum_c \frac{\left| \sum_\alpha \sum_{b\text{ for } {\{b,c\} \in \alpha}} \data_{bc}^* \hat{\gains}_b \hat{\fitparamsu}_\alpha \right|}{\sum_\alpha \sum_{b\text{ for } {\{b,c\} \in \alpha}} |\hat{\gains}_b \hat{\fitparamsu}_\alpha|^2}.
\end{equation}
We have a further extremum condition from the fit visibilities $\fitparamsu$:
\begin{equation}
    \left. \frac{\partial \negloglikelihood(\gains, \fitparamsu)}{\partial \fitparamsu_\alpha} \right|_{\gains=\hat{\gains}, \fitparamsu=\hat{\fitparamsu}} = 0.
\end{equation}
Solving gives
\begin{equation}
    \hat{\fitparamsu}_\alpha = \frac{\frac{1}{\sigma_\text{T}^2} \sum_{\{b,c\} \in \alpha} \data_{bc} \hat{\gains}_b^* \hat{\gains}_c + \frac{1}{\sigma_\text{M}^2} \modelvals_\alpha}{\frac{1}{\sigma_\text{T}^2} \sum_{\{b,c\} \in \alpha} |\hat{\gains}_b \hat{\gains}_c^*|^2 + \frac{1}{\sigma_\text{M}^2}}.
\end{equation}

If we again let $\gains = A \unnormgains$ and take the first-order approximation $\hat{\unnormgains} \approx 1$, we get
\begin{equation}
    \hat{\gainamp}^2 \approx \frac{1}{N_\text{ant}} \sum_c \frac{\left| \sum_\alpha \sum_{b\text{ for } {\{b,c\} \in \alpha}} \data_{bc}^* \hat{\fitparamsu}_\alpha \right|}{\sum_\alpha |\hat{\fitparamsu}_\alpha|^2} \approx \frac{\left| \langle \data^* \hat{\fitparamsu} \rangle \right|}{ \langle |\hat{\fitparamsu}|^2 \rangle}
\label{eq:unified_cal_approx_gain_amp}
\end{equation}
and
\begin{equation}
    \hat{\fitparamsu}_\alpha \approx \frac{\frac{1}{\sigma_\text{T}^2} \hat{\gainamp}^2 \sum_{\{b,c\} \in \alpha} \data_{bc} + \frac{1}{\sigma_\text{M}^2} \modelvals_\alpha}{\frac{1}{\sigma_\text{T}^2} \hat{\gainamp}^4 N_\alpha + \frac{1}{\sigma_\text{M}^2}},
\label{eq:unified_cal_approx_fit_vis}
\end{equation}
where $N_\alpha$ is the number of baselines in redundant set $\alpha$.

We can explore the opposing limits that $\sigma_\text{M} \rightarrow 0$ and $\sigma_\text{M} \rightarrow \infty$. For $\sigma_\text{M} \rightarrow 0$, Equation \ref{eq:unified_cal_approx_fit_vis} gives $\hat{\fitparamsu}_\alpha \approx \modelvals_\alpha$. We then get that
\begin{equation}
    \hat{\gainamp}^2 \approx \frac{\left| \langle \data^* \modelvals \rangle \right|}{ \langle |\modelvals|^2 \rangle},
\end{equation}
which is identical to the solution for sky-based calibration in \S\ref{s:sky_cal_gain_amp}. In the opposite limit that $\sigma_\text{M} \rightarrow \infty$, Equation \ref{eq:unified_cal_approx_fit_vis} becomes
\begin{equation}
    \hat{\fitparamsu}_\alpha \approx \frac{  \sum_{\{b,c\} \in \alpha} \data_{bc}}{ \hat{\gainamp}^2 N_\alpha}.
\end{equation}
Following Equation \ref{eq:data_decomp} and again assume no correlation between $\boldsymbol{n}_\text{T}$ and $\fitparamsu_\text{true}$, we find that 
\begin{equation}
    \left| \langle \data^* \hat{\fitparamsu} \rangle \right| \approx \frac{1}{\hat{\gainamp}^2} \left( \langle |\fitparamsu_\text{true}|^2 \rangle + \langle |\boldsymbol{n}_\text{T}|^2 \rangle \right)
\end{equation}
and
\begin{equation}
    \langle |\hat{\fitparamsu}|^2 \rangle \approx \frac{1}{\hat{\gainamp}^4} \left( \langle |\fitparamsu_\text{true}|^2 \rangle + \langle |\boldsymbol{n}_\text{T}|^2 \rangle \right).
\end{equation}
We therefore get that $\hat{\gainamp}^2 \approx \hat{\gainamp}^2$, reflecting that the overall amplitudes of the gains is degenerate in the relative calibration step of redundant calibration.

In unified calibration $\sigma_\text{M}$ is finite and nonzero. In order to evaluate the average gain amplitude, we make the approximation that $N_\alpha$ is constant across redundant baseline sets. This is a non-physical assumption---for the simulation in \S\ref{s:sim_results} the number of baselines in each redundant set varies from 1 to 30---but is useful for simplifying Equation \ref{eq:unified_cal_approx_gain_amp} and deriving a rough approximation of the average gain amplitude in unified calibration. Under this approximation, we evaluate
\begin{equation}
    \left| \langle \data^* \hat{\fitparamsu} \rangle \right| 
    \approx \frac{\left| \frac{1}{\sigma_\text{T}^2} \hat{\gainamp}^2 \langle N \rangle \langle |\data|^2 \rangle+ \frac{1}{\sigma_\text{M}^2} \langle \data^* \modelvals \rangle \right|}{\frac{1}{\sigma_\text{T}^2} \hat{\gainamp}^4 \langle N \rangle+ \frac{1}{\sigma_\text{M}^2}} 
    \approx \frac{ \left(\frac{1}{\sigma_\text{T}^2} \hat{\gainamp}^2 \langle N \rangle + \frac{1}{\sigma_\text{M}^2} \right) \langle |\fitparamsu_\text{true}|^2 \rangle+  \frac{1}{\sigma_\text{T}^2} \hat{\gainamp}^2 \langle N \rangle \langle |\boldsymbol{n}_\text{T}|^2 \rangle }{\frac{1}{\sigma_\text{T}^2} \hat{\gainamp}^4 \langle N \rangle+ \frac{1}{\sigma_\text{M}^2}}
\end{equation}
and
\begin{equation}
\begin{split}
    \langle |\hat{\fitparamsu}|^2 \rangle 
    &\approx \frac{ \frac{1}{\sigma_\text{T}^4} \hat{\gainamp}^4 \langle N \rangle^2 \langle |\data|^2 \rangle+ \frac{1}{\sigma_\text{M}^4} \langle |\modelvals|^2 \rangle + \frac{2}{\sigma_\text{T}^2 \sigma_\text{M}^2} \hat{\gainamp}^2 \langle N \rangle \langle \operatorname{Re}(\data^*\modelvals)\rangle}{\left(\frac{1}{\sigma_\text{T}^2} \hat{\gainamp}^4 \langle N \rangle+ \frac{1}{\sigma_\text{M}^2}\right)^2} \\
    &\approx \frac{ 
    \left(\frac{1}{\sigma_\text{T}^2} \hat{\gainamp}^2 \langle N \rangle + \frac{1}{\sigma_\text{M}^2} \right)^2 \langle |\fitparamsu_\text{true}|^2 \rangle + \frac{1}{\sigma_\text{T}^4} \hat{\gainamp}^4 \langle N \rangle^2 \langle |\boldsymbol{n}_\text{T}|^2 \rangle + \frac{1}{\sigma_\text{M}^4} \langle |\boldsymbol{n}_\text{M}|^2 \rangle 
    }{\left(\frac{1}{\sigma_\text{T}^2} \hat{\gainamp}^4 \langle N \rangle+ \frac{1}{\sigma_\text{M}^2}\right)^2},
\end{split}
\end{equation}
where $\langle N \rangle$ is the average number of baselines in a redundant set. Combining these, we get
\begin{equation}
    \hat{\gainamp}^2 \approx \left( \frac{1}{\sigma_\text{T}^2} \hat{\gainamp}^4 \langle N \rangle+ \frac{1}{\sigma_\text{M}^2} \right)
    \frac{
    \left(\frac{1}{\sigma_\text{T}^2} \hat{\gainamp}^2 \langle N \rangle+ \frac{1}{\sigma_\text{M}^2} \right) \langle |\fitparamsu_\text{true}|^2 \rangle+  \frac{1}{\sigma_\text{T}^2} \hat{\gainamp}^2 \langle N \rangle \langle |\boldsymbol{n}_\text{T}|^2 \rangle
    }{
    \left(\frac{1}{\sigma_\text{T}^2} \hat{\gainamp}^2 \langle N \rangle+ \frac{1}{\sigma_\text{M}^2} \right)^2 \langle |\fitparamsu_\text{true}|^2 \rangle + \frac{1}{\sigma_\text{T}^4} \hat{\gainamp}^4 \langle N \rangle^2 \langle |\boldsymbol{n}_\text{T}|^2 \rangle + \frac{1}{\sigma_\text{M}^4} \langle |\boldsymbol{n}_\text{M}|^2 \rangle
    },
\end{equation}
which can be solved numerically to evaluate $\hat{\gainamp}$.

As noted in \S\ref{s:sky_cal_gain_amp}, in the simulation presented in \S\ref{s:random_error_sim} we have $\left\langle |\fitparamsu_\text{true}|^2 \right\rangle = 38.45 \text{ Jy}^2$ and $\left\langle |\boldsymbol{n}_\text{M}|^2 \right\rangle = 0.31 \text{ Jy}^2$. We also have $\left\langle |\boldsymbol{n}_\text{T}|^2 \right\rangle = 0.08 \text{ Jy}^2$, $\sigma_\text{M} = 0.4 \text{ Jy}$, $\sigma_\text{T} = 0.2 \text{ Jy}$, and $\langle N \rangle = 10.5$. Plugging in these values and numerically solving, we get $\hat{\gainamp} \approx 1.0009$. In the simulation presented in \S\ref{s:random_error_sim}, unified calibration returns an average gain amplitude of 1.0005.

\end{document}